\documentclass[aps,prx,twocolumn,longbibliography,groupedaddress]{revtex4-2} 

\usepackage{epsfig}
\usepackage{subfigure}
\usepackage{graphicx}
\usepackage{amsfonts}
\usepackage[figuresright]{rotating}
\usepackage{amssymb}
\usepackage{amsmath}
\usepackage{psfrag}
\usepackage{esint} 
\usepackage{bm}
\usepackage[colorlinks,linkcolor=blue,anchorcolor=blue,citecolor=blue,urlcolor=blue]{hyperref}
\usepackage[version=4]{mhchem}

\newtheorem{theorem}{Theorem}

\newenvironment{proof}[1][Proof]{\noindent \textbf{#1: }}{\ \rule{0.5em}{0.5em}}

\def\be{\begin{equation}} \def\ee{\end{equation}}
\def\bea{\begin{eqnarray}} \def\eea{\end{eqnarray}}

\def\k{{\bf k}}

\def\r{{\bf r}}
\def\R{{\bf R}}

\def\v{{\bf v}}
\def\u{{\bf u}}
\def\w{{\bf w}}

\renewcommand{\vec}[1]{\mathbf{#1}}
\newcommand{\vecg}[1]{\boldsymbol{#1}}

\def\bpm{\begin{pmatrix}} \def\epm{\end{pmatrix}}

\newcommand{\pxup}{p_{\r, \hat{x},\uparrow}}
\newcommand{\pxdown}{p_{\r, \hat{x},\downarrow}}
\newcommand{\pyup}{p_{\r, \hat{y},\uparrow}}
\newcommand{\pydown}{p_{\r, \hat{y},\downarrow}}

\makeatletter
\newcommand*{\balancecolsandclearpage}{%
  \close@column@grid
  \clearpage
}
\makeatother

\begin{document}
\title{Hund's coupling-assisted ferromagnetic percolation transition in a multiorbital \\ flat band}

\author{Eric Bobrow}
\author{Junjia Zhang}
\author{Yi Li}
\affiliation{Department of Physics and Astronomy, Johns Hopkins University, Baltimore, Maryland 21218, USA}
\date{August 26, 2021}

\begin{abstract}
    By connecting Hund's physics with flat band physics, we establish an exact result for studying ferromagnetism in a multiorbital system. 
    We consider a two-layer model consisting of a $p_x$, $p_y$-orbital honeycomb lattice layer and an $f$-orbital triangular lattice layer with sites aligned with the centers of the honeycomb plaquettes. The system features a flat band that admits a percolation representation for an appropriate chemical potential difference between the two layers. In this representation, the ground state space is spanned by maximum-spin clusters of localized single-particle states, and averaging over the ground states yields a correlated percolation problem with weights due to the spin degeneracy of the clusters. A paramagnetic-ferromagnetic transition occurs as the band approaches half filling and the ground states become dominated by states with a large maximum-spin cluster, as shown by Monte Carlo simulation.
\end{abstract}
\maketitle

\section{Introduction}
Flat band physics gives rise to rich phases of matter in the presence of interactions, ranging from Mott insulating and Wigner crystal states to magnetism and fractional quantum Hall type topological states  
\cite{Mielke1993a, Mielke1991, Wu2007, Tang2011, Bistritzer2011, Wang2013, Po2018, Tarnopolsky2019, Liu2019b, Bergman2008, Creutz2001, Zurita2020, Chiu2020, Bergholtz2013, Liu2014FlatBand, Derzhko2015}.
However, reaching a rigorous theoretical understanding of many-body states in flat bands is difficult, since flat bands enhance interaction effects. As a result, nonperturbative techniques and exact results are particularly valuable. 

Flat bands can often be attributed to a large number of degenerate states that are spatially localized due to destructive interference. 
Tight-binding models on line graphs \cite{Mielke1991} as well as certain decorated lattices such as the Tasaki lattice \cite{Tasaki1992} feature flat bands that can be understood through destructively interfering hopping, as can the flat bands in many more general graphs \cite{Tanaka2020}. 
In the presence of a repulsive on-site Hubbard interaction, these models exhibit saturated ferromagnetism when the flat band is half filled \cite{Mielke1991a, Mielke1993a}. 
Certain models such as the Tasaki lattice possess a provable ferromagnetic phase extending below half filling by realizing a direct exchange mechanism between overlapping localized states. 
In these models, interacting ground states are degenerate and spanned by different configurations of clusters of localized states, with the clusters independently maximizing spin \cite{Mielke1992,Mielke1993a}. 
Since the total spin of each cluster depends only on its size, the result is a correlated \textit{percolation representation} that can be efficiently simulated to find the transition from paramagnetic states with small clusters at low filling to ferromagnetic states with large clusters at high filling \cite{Maksymenko2012,Liu2019a}.

Orbital degrees of freedom, essential in many real materials, can be vital to the formation of a flat band, as orbital-dependent anisotropic hopping can facilitate the formation of localized states. A particular example of this is the honeycomb lattice with $p_x$ and $p_y$ orbitals at each site and orbital-dependent nearest neighbor hopping along the bond directions, where it was shown in Ref. [\onlinecite{Wu2007}] that both the lowest and highest bands are flat.
The flat band degeneracy can be attributed to loop states localized around each honeycomb plaquette where the $p$ orbital at each site is oriented perpendicularly to the outgoing bond, preventing hopping out of the plaquette due to the bond-projected hopping. 

However, in the presence of interactions, flat band localized states do not guarantee a provable ferromagnetic transition based on the percolation representation. 
Unlike in the Tasaki lattice, repulsive Hubbard interactions do not lead to an immediate percolation representation in the $p_x$, $p_y$-orbital honeycomb model. 
To take advantage of the direct exchange mechanism, Mielke and Tasaki's scheme \cite{Mielke1993a} requires that no more than two flat band localized states overlap at any site and that each state satisfy a quasilocality condition, which leads to a preference for each flat band localized state to be at most singly occupied.
In multiorbital systems, the conditions in Mielke and Tasaki's scheme are often violated, which can break the percolation representation. 
For the $p$-orbital honeycomb model \cite{Wu2007}, the loop states violate both conditions, with three loop states overlapping at each $p_y$ orbital on each site. This can introduce additional states into the interacting ground state space that have lower spin within the cluster.

In this article, we show that the combination of Hund's physics and flat band physics can allow a percolation representation to be found for multiorbital systems. 
We construct a two-layer model consisting of a $p_x$- and $p_y$-orbital honeycomb layer together with a triangular lattice $f$-orbital layer. For appropriate chemical potential difference between the layers, the system features a flat band spanned by localized states centered on each $f$ orbital. We show that when this band is the highest energy band and at least half filled, the system admits a percolation representation in the presence of Hund's coupling between the $p$ orbitals. 
This result goes beyond Mielke and Tasaki's scheme \cite{Mielke1993a} and allows three localized states to overlap on site, which is useful for honeycomb lattice systems.

The remainder of this article is organized as follows. 
In Section \ref{sec:model}, we introduce the multiorbital flat band model with intraorbital Hubbard interactions and Hund's coupling. We then discuss the localized single-particle flat band states that will be useful for establishing the percolation representation.
In Section \ref{sec:perc}, we first review Mielke and Tasaki's scheme for percolation representations, then prove a Hund's coupling-assisted percolation representation using the localized flat band states for our multiorbital system, and finally present Monte Carlo simulation results for the ferromagnetic percolation transition.
\section{Multiorbital Flat Band Model}
\label{sec:model}

In order to find a multiorbital flat band system that admits a percolation representation, we construct a model where the flat band is described by suitable localized states. 
The model system consists of spin-$1/2$ electrons in a two-layer system with one layer a $p_x$, $p_y$-orbital honeycomb lattice and the other a triangular lattice with one $f_{y(3x^2-y^2)}$ orbital per site. The $f$-orbital triangular lattice sites are aligned with the centers of the $p$-orbital honeycomb plaquettes. 
The set of p-orbital and f-orbital sites will respectively be labeled $\Lambda_p$ and $\Lambda_f$, with the overall two-layer lattice $\Lambda = \Lambda_p \sqcup \Lambda_f$. 

\begin{figure}[tbp]
\subfigure[]{\epsfig{file=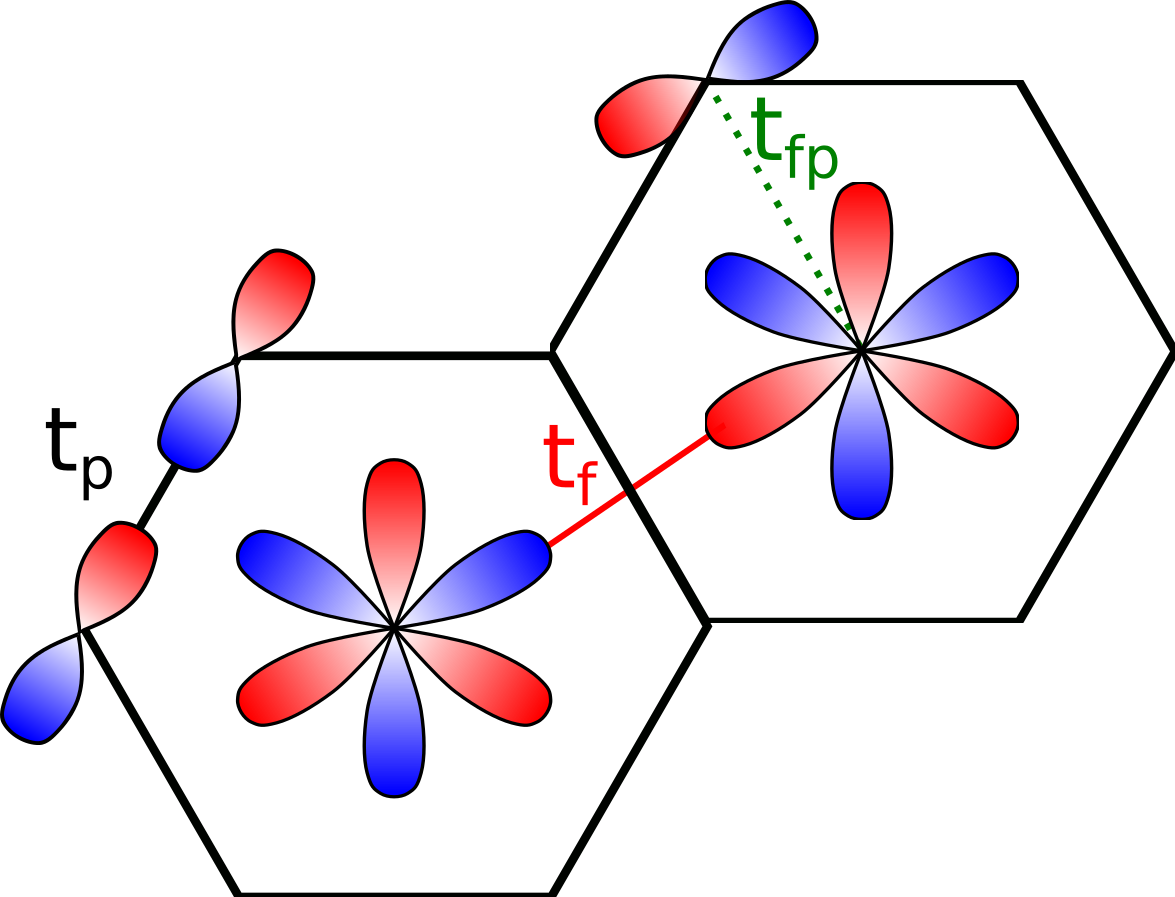, width=0.38\linewidth}
\label{fig:hop}}
\subfigure[]{\epsfig{file=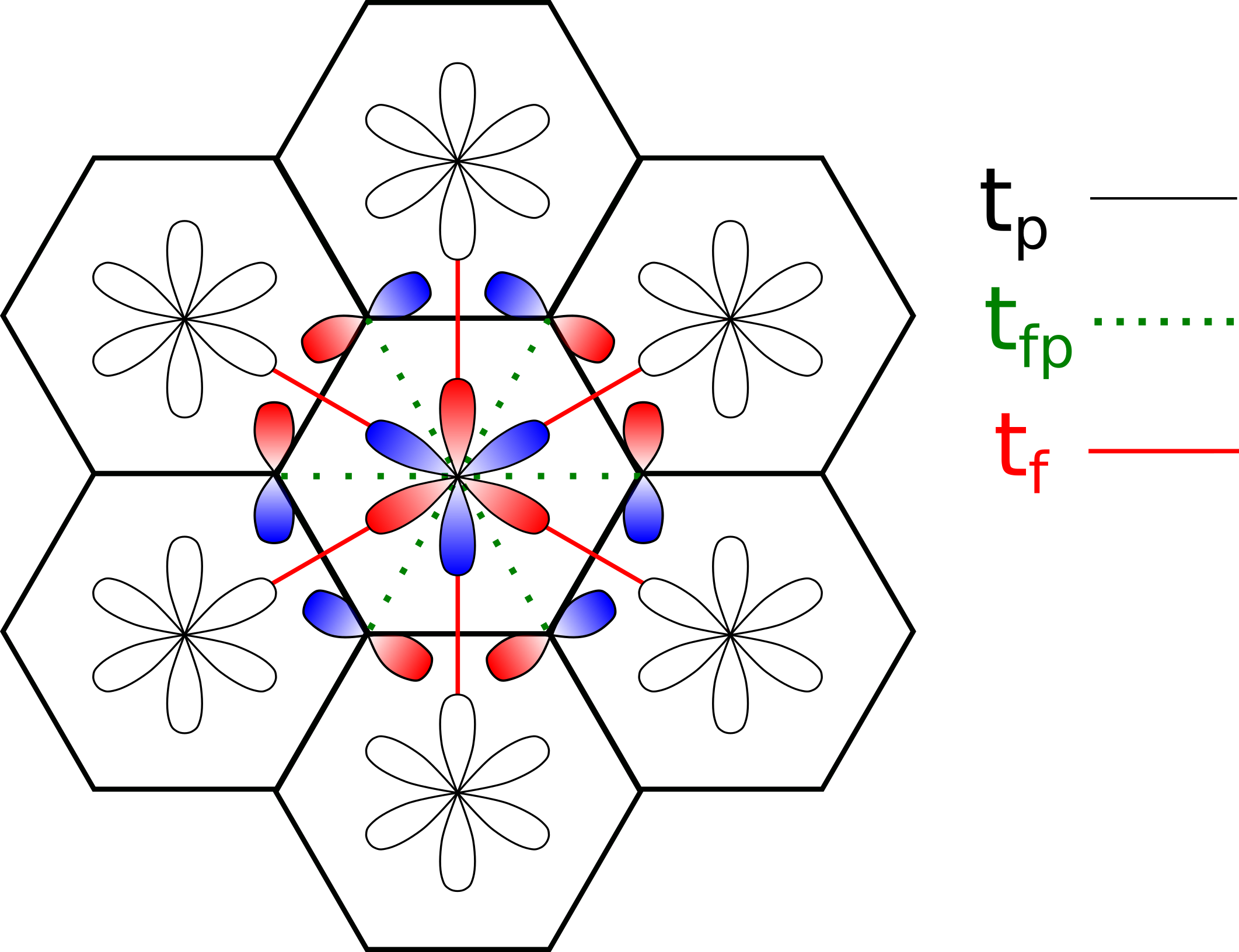, width=0.58\linewidth}
\label{fig:state}}
\caption{Top view of the two-layer lattice. Red and blue respectively denote positive and negative lobes for the $p$- and $f$-orbital wavefunctions. (a) Examples are shown of hopping $t_f$ between $f$ orbitals, $t_p$ between $p$ orbitals along the honeycomb bond direction, and $t_{fp}$ between $f$ and $p$ orbitals perpendicular to the hopping direction. (b) A localized state formed by a superposition of an $f$ orbital and a surrounding loop of $p$ orbitals with alternating signs.}
\label{fig:model}
\end{figure}

For our model, the kinetic part of the Hamiltonian includes hopping within the $p$- and $f$-orbital layers as well as between nearest $f$ and $p$ orbitals, $H_K = H_K^p + H_K^f + H_K^{fp}$, with hopping terms depicted in Fig. \ref{fig:hop}. 
$H_K^p$ consists of nearest neighbor hopping between $p$ orbitals projected along the honeycomb bond direction, which describes $\sigma$ bonding. 
\begin{equation}
    H_K^p = t_p \sum_{\r\in \Lambda_{p}^A} \sum_{\sigma = \uparrow,\downarrow} \sum_{i = 1}^3 p^\dagger_{\r + \v_i, \v_i, \sigma} p_{\r, \v_i, \sigma} + h.c.,
\label{eq:hp}
\end{equation}
where the hopping amplitude $t_p > 0$, since the projected $p$ orbitals are odd under reflection through the $\sigma$ bond. $\Lambda_p^A$ is the $A$ sublattice of the $p$-orbital layer and $p_{\r, \v_i, \sigma} = \hat{v}_i \cdot \vec{p}_{\r, \sigma}$ is the projection of the $p_x$, $p_y$ orbitals at site $\r$ in the bond direction $\hat{\v}_i$. The vectors $\v_i$ are the nearest neighbor vectors $v_1 = (1,0)$, $v_2= (-\frac{1}{2}, \frac{\sqrt{3}}{2})$, $v_3 = (-\frac{1}{2}, -\frac{\sqrt{3}}{2})$. Here we take the honeycomb lattice bond length to be $1$. 
This $p_x$, $p_y$-orbital model has been previously discussed, for example, in Refs. [\onlinecite{Wu2007, Wu2008c, Zhang2010a, Zeng2021}].

The $f$ orbital layer forms a triangular lattice with sites $\R$ aligned with the centers of the $p$-orbital honeycombs and features nearest-neighbor hopping given by
\begin{equation}
    H_K^f = t_f \sum_{\R \in \Lambda_f} \sum_{\sigma = \uparrow, \downarrow} \sum_{i = 1}^6 f^\dagger_{\R + \w_i, \sigma} f_{\R, \sigma} + h.c.,
\label{eq:hf}
\end{equation}
where $t_f > 0$, for reasons similar to $t_p > 0$, and $\w_i$ are the six nearest-neighbor vectors on the $f$-orbital triangular lattice, $w_i = \sqrt{3}(\cos \phi_i, \sin\phi_i)$ with $\phi_i = \frac{\pi}{6}(2i - 1)$ and $i=1, \cdots, 6$.

Hopping between layers involves nearest $p$- and $f$-orbital sites. Due to symmetry of the $f_{y(3x^2-y^2)}$ orbitals, only the $p_x,p_y$ orbitals directed perpendicularly to the in-plane hopping direction are involved in the hopping, as shown in Fig. \ref{fig:hop}. The hopping between the layers is then described by
\begin{equation}
    H_K^{fp} = t_{fp} \sum_{\R\in\Lambda_f} \sum_{\sigma=\uparrow, \downarrow} \sum_{i = 1}^6 (-1)^{i-1}f^\dagger_{\R, \sigma} p_{\R+\u_i, \u^\perp_i, \sigma} + h.c.,
\label{eq:hfp}
\end{equation}
where $t_{fp} > 0$ and $\u_i = (\cos\theta_i, \sin\theta_i)$ 
with $\theta_i = \frac{\pi}{3} (i-1)$, a reordering of the $\v$ vectors. In particular, $\R + \u_i \in \Lambda_p$ when $\R\in \Lambda_f$.  Here $\u^\perp_i$ are defined to be unit vectors perpendicular to $\u_i$ with sign chosen so that $\u^\perp_{i+1}$ is a $\pi/3$ rotation of $\u_i$. Thus, $\u^\perp_i = (-\sin \theta_i, \cos \theta_i)$. The alternating sign in the hopping amplitude is due to the fact that the $f_{y(3x^2-y^2)}$ orbital changes sign under $\frac{\pi}{3}$ rotation. To interpret the model as a two-layer system, the $\u_i$ vectors can be thought of as the $xy$-plane component of the vector between the $f$ orbitals on $\Lambda_f$ and the adjacent $p$ orbitals on $\Lambda_p$, with the $z$ component being a small interlayer distance.

The chemical potentials for the $p$ and $f$ orbitals are given by $H_\mu = -\mu_p \hat{N}_p - \mu_f \hat{N}_f$, where the total number operator for $p$-orbital electrons is $\hat{N}_p = \sum_{\r\in\Lambda_p}\sum_{p=p_x, p_y} \sum_{\sigma=\uparrow\downarrow}n_{\r, p, \sigma}$ with $n_{\r, p_x,\sigma} = p^\dagger_{\r, \hat{x},\sigma} p_{\r, \hat{x}, \sigma}$ and a similar expression for $n_{\r, p_y,\sigma}$. 
The total number operator for $f$-orbital electrons is defined similarly, $\hat{N}_f = \sum_{\R\in\Lambda_f} \sum_{\sigma=\uparrow\downarrow}n_{\R, f, \sigma}$ with $n_{\R, f, \sigma} = f^\dagger_{\R, \sigma} f_{\R, \sigma}$.
In a layered material, the chemical potential difference between the $p$- and $f$-orbital layers can in principle be controlled by gating. 
For an appropriate chemical potential difference, we will see that the model admits the desired flat band for any particular hopping amplitudes.

The on-site intraorbital Coulomb interaction is described by the Hubbard $U$ terms for $p$ and $f$ orbitals. Written in a particle-hole-symmetric form, the Hubbard interaction Hamiltonian $H_U$ is 
\begin{equation}
\begin{aligned}
    H_U &= U_p \sum_{\r\in\Lambda_p} \sum_{p = p_x, p_y} \left(n_{\r, p, \uparrow} -\frac{1}{2}\right)\left(n_{\r, p, \downarrow}-\frac{1}{2}\right) \\
    &+ U_f\sum_{\R\in\Lambda_f} \left(n_{\R, f, \uparrow}-\frac{1}{2}\right)\left(n_{\R, f, \downarrow}-\frac{1}{2}\right),
\end{aligned}
\label{eq:hu}
\end{equation}
which can also be written in terms of $H_U' = U_p \sum_{\r\in\Lambda_p}\sum_{p=p_x,p_y} n_{\r, p, \uparrow}n_{\r,p,\downarrow} + U_f\sum_{\R\in\Lambda_f}n_{\R,f,\uparrow}n_{\R,f,\downarrow}$ as $H_U = H_U' -\frac{U_p}{2}\hat{N}_p - \frac{U_f}{2}\hat{N}_f + \frac{2U_p|\Lambda_p| + U_f|\Lambda_f|}{4}$. This alternate expression, which separates the two-particle interaction $H_U'$ from terms that shift the chemical potential will be convenient for studying the interacting ground states in Sec. \ref{sec:perc}.
We will consider only the repulsive case with $U_p, U_f > 0$.

As we will see, the intraorbital Hubbard interactions alone are not sufficient for the percolation representation we discuss in Section. \ref{sec:fmodelperc}. The ability of the Hund's coupling in multiorbital systems to polarize electrons in degenerate orbitals will be essential. The on-site Hund's coupling between $p_x$ and $p_y$ orbitals is
\begin{equation}
    H_J = -J \sum_{\r\in\Lambda_p} \left(\vec{S}_{\r, p_x} \cdot \vec{S}_{\r, p_y}-\frac{1}{4} n_{\r, p_x} n_{\r, p_y}\right),
\label{eq:hj}
\end{equation}
where $S^i_{\r, p_{x/y}} = \frac{1}{2}\sum_{\mu, \nu = \uparrow, \downarrow} p^\dagger_{\r,\hat{x}/\hat{y},\mu} \sigma^i_{\mu\nu} p_{\r, \hat{x}/\hat{y}, \nu}$ with Pauli matrices $\sigma^i$, $i=x,y,z$, and the Hund's coupling $J>0$. The number operators without a spin index count both spin up and down, such as $n_{\r, p_x} =\sum_{\sigma=\uparrow, \downarrow}n_{\r, p_x, \sigma}$. 
The Hund's coupling energy is $J$ for an on-site interorbital singlet and zero for a triplet state.

We now study the structure of the flat bands in the noninteracting model and examine the localized flat band states.
In the absence of the hopping between $f$ and $p$ orbitals, 
the $p$ orbital layer described by $H_K^p - \mu_p \hat{N}_p$ is known to exhibit flat bands at energies $E_{p,\pm} = \pm \frac{3}{2}t_p-\mu_p$ \cite{Wu2007, Wu2008c, Zhang2010a}. 
The lower flat band for spin $\sigma$ is spanned by localized $p$-orbital loops with alternating sign $|\psi^{-(p)}_{\R, \sigma}\rangle = \frac{1}{\sqrt{6}}\sum_{i = 1}^6 (-1)^{i-1} p^\dagger_{\R + \u_i, \u_i^\perp,  \sigma}|0\rangle$ on each honeycomb, where the $f$ orbital site $\R\in\Lambda_f$ is used to label the surrounding honeycomb.
The upper flat band is spanned by $|\psi^{+(p)}_{\R, \sigma}\rangle = \frac{1}{\sqrt{6}}\sum_{i = 1}^6 p^\dagger_{\R + \u_i, \u_i^\perp, \sigma}|0\rangle$. These loop states feature a superposition of $p_x$ and $p_y$ orbitals to form a $p$ orbital perpendicular to the outgoing bond that cannot hop out due to the bond-projected hopping. These states can also be thought of as localized due to destructive interference in the $p_x$, $p_y$ basis.

In the presence of the $f$-orbital layer and interlayer hopping, $H_{fp}$ hybridizes the $p$- and $f$-orbital bands. In general, this hybridization disperses the $p$-orbital flat band at energy $E_{p,-}$, while the flat band at $E_{p, +}$ remains flat due to destructive interference at the $f$ orbital sites. 
We can find conditions under which there is a flat band at an energy $E_{fp}$ with eigenstates involving both $p$ and $f$ orbitals by taking the ansatz $|\psi_{\R, \sigma}\rangle \equiv a|\psi^{-(p)}_{\R,\sigma}\rangle + b |f_{\R, \sigma}\rangle$, with $a$ and $b$ determined by
\begin{equation}
    \begin{aligned}
    H_K |\psi_{\R, \sigma}\rangle &= (E_{p,-}-\mu_p) a|\psi^{-(p)}_{\R,\sigma}\rangle + 6t_{fp} a |f_{\R,\sigma}\rangle \\
    & +t_{fp} b|\psi^{-(p)}_{\R,\sigma}\rangle -\mu_f b |f_{\R, \sigma}\rangle \\
    &+ (-a t_{fp}  + b t_f)\sum_{i=1}^6 |f_{\R + \w_i,\sigma}\rangle\\
    &= E_{fp}|\psi_{\R, \sigma}\rangle.
    \end{aligned}
\end{equation}
The state 
\begin{equation}
    |\psi_{\R, \sigma}\rangle = 
{\mathcal{N}_t} (t_f|\psi^{-(p)}_{\R,\sigma}\rangle +  t_{fp}|f_{\R,\sigma}\rangle)
\label{eq:psi}
\end{equation}
with normalization $\mathcal{N}_t =1/\sqrt{t_f^2+t_{fp}^2}$
is an eigenstate with energy $E_{fp} = E_{p,-} + t_{fp}^2/t_f$ as long as $\mu_f = \mu_f^c$ with
\begin{equation}
    \mu_f^c \equiv  \mu_p + 6 t_f +\frac{3}{2}t_p - \frac{t_{fp}^2}{t_f}.
    \label{eq:mufc}
\end{equation}
The state $|\psi_{\R, \sigma}\rangle$ is localized to the $f$ orbital site $\R$ and the neighboring $p$ orbital sites, as shown in Fig. \ref{fig:state}. These localized states at different $\R$ are degenerate and form a complete basis for the flat band with energy $E_{fp}$. 
To see that $|\psi_{\R, \sigma}\rangle$ for all $\R$ span the $E_{fp}$ flat band, note that these states are linearly independent since only $|\psi_{\R, \sigma}\rangle$ has nonzero amplitude at the $f$ orbital site $\R$. This set of linearly independent states has the same dimension as the flat band, the $E_{fp}$ eigenspace, since one state is associated with each unit cell. This argument assumes no other bands touch the $E_{fp}$ flat band, which is true for a range of parameters as we will now see by examining the dispersive bands.

\begin{figure}[tbp]
\subfigure[]{\epsfig{file=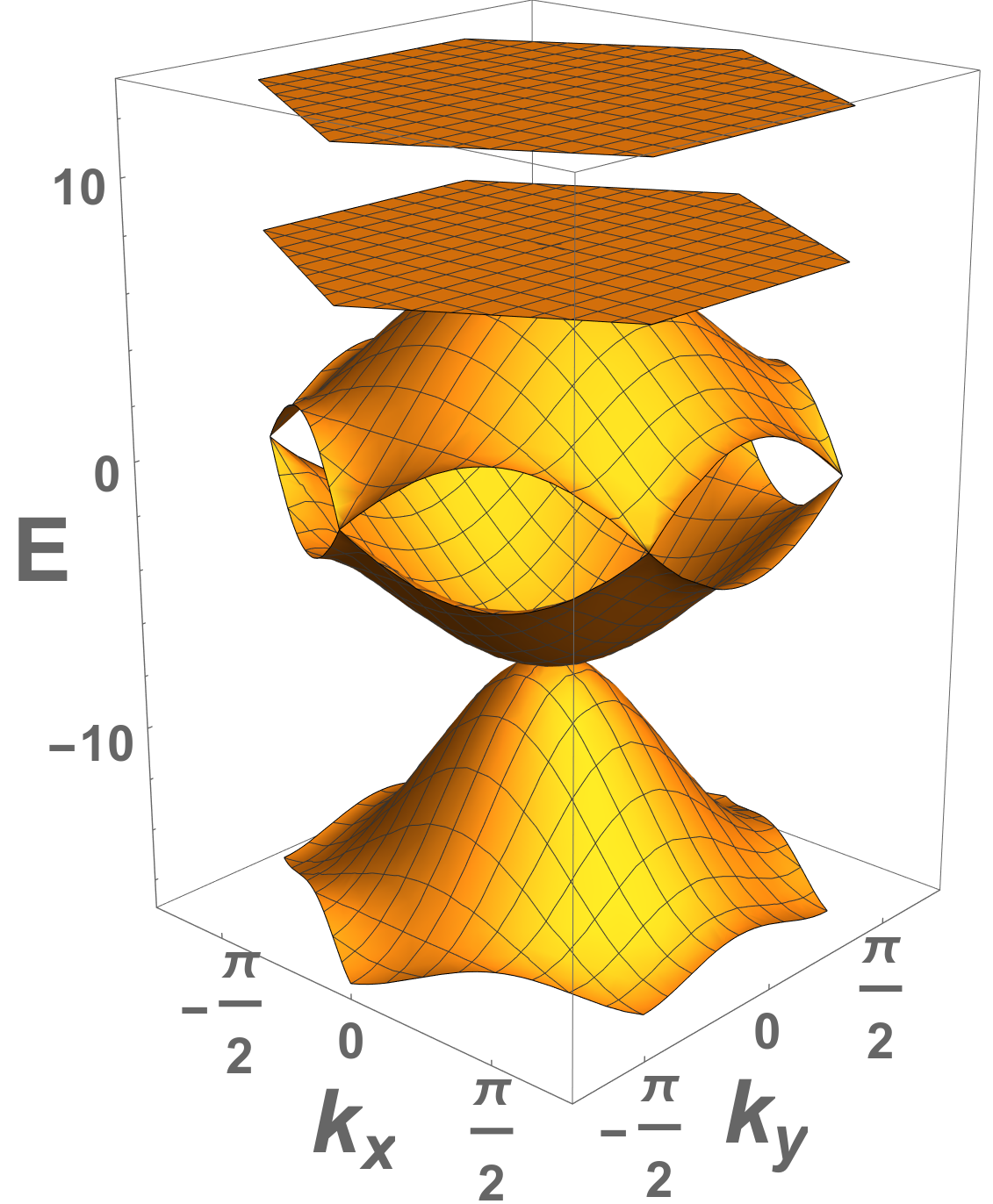, width=0.38\linewidth}
\label{fig:spect}}
\subfigure[]{\epsfig{file=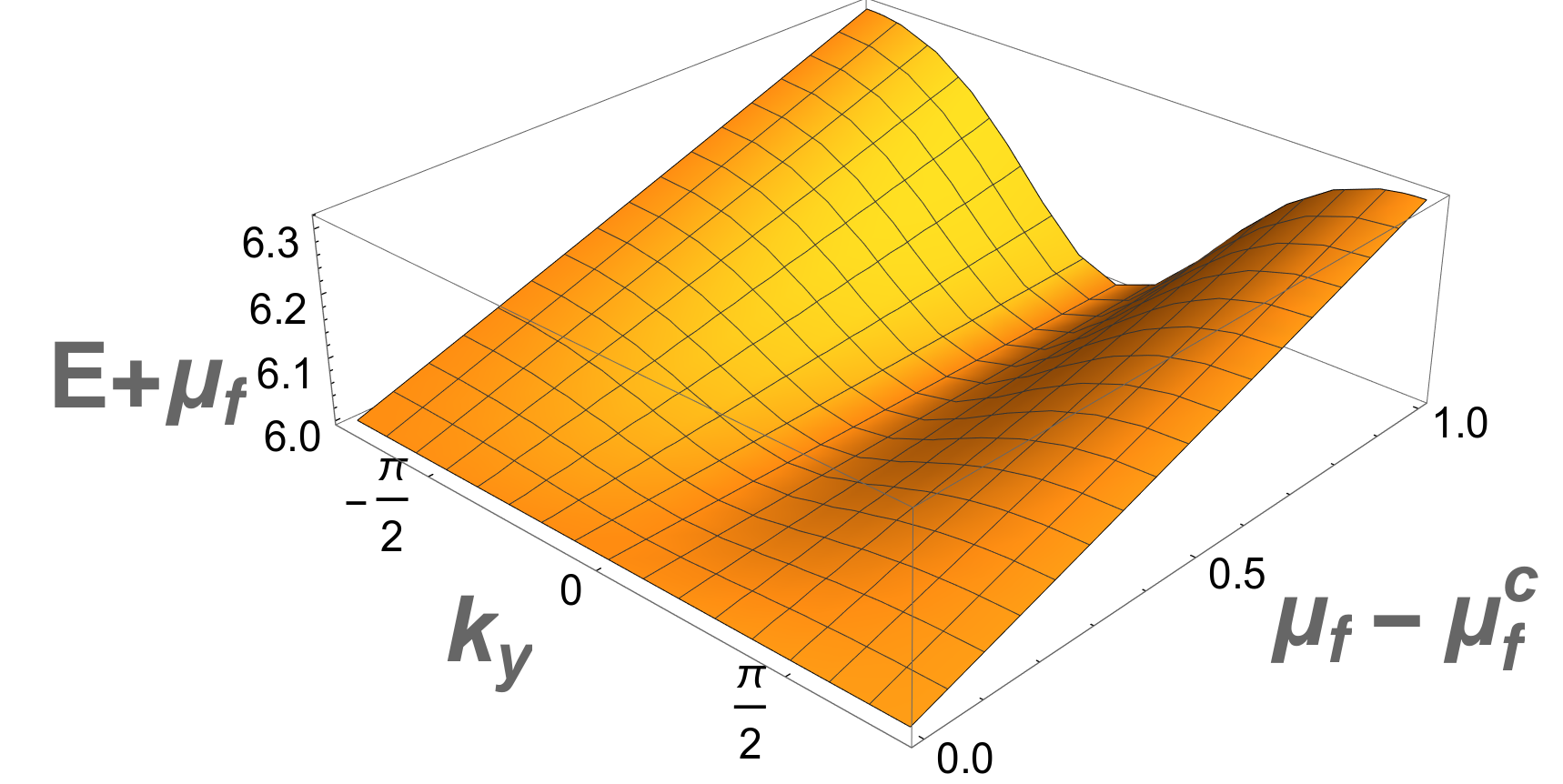, width=0.58\linewidth}
\label{fig:curv}}
\caption{For $\mu_p = 0$, $t_p = 5$, $t_f = 1$, and $t_{fp} = 4.5$, plots are shown of (a) the band structure of $H_K$ in the first Brillouin zone with $\mu_f = \mu_f^c$ and (b) the energy of the top band as a function of $k_y$ and $\mu_f$. At $\mu_f = \mu_f^c$, the top band is flat.}
\label{fig:spectplots}
\end{figure}

When $\mu_f = \mu_f^c$, there are five bands including the two flat bands at energies $E_{fp}$ and $E_{p,+}$ and three dispersive bands with energies 
\begin{equation}
    \begin{aligned}
    E_{2, \pm}(\k) &= -\mu_p \pm \frac{t_p}{2}\sqrt{4\cos{\frac{3 k_x}{2}} \cos{\frac{\sqrt{3} k_y}{2}} + 2\cos{\sqrt{3}k_y} + 3}, \\
    E_{1}(\k) &= -\frac{3 t_p}{2} - \mu_p \\
    &+ 2 t_f \left(2\cos{\frac{3 k_x}{2}} \cos{\frac{\sqrt{3} k_y}{2}} + \cos{\sqrt{3}k_y} - 3\right),
    \end{aligned}
\end{equation}
as shown in Fig. \ref{fig:spect}.
In fact, $E_{2, \pm}$ are exactly the dispersive bands of the $p$-orbital model $H^p_K$ \cite{Wu2007}. When $\mu = \mu_f^c$, the addition of the $f$-orbital layer leaves the dispersive bands and one flat band, $E_{p,+}$, of the $p$-orbital layer unchanged, while introducing a dispersive band $E_1(\k)$ and involving $f$ orbitals in the remaining flat band $E_{fp}$.
The flat band at energy $E_{fp}$ will be useful for the percolation representation. It is the highest energy band and well separated from the lower bands when $t_{fp}^2 > 3t_{p}t_f$, which we will assume is the case below. 
In Fig. \ref{fig:curv}, the energy of the top band with a $\mu_f$ shift is plotted as a function of $\mu_f - \mu_f^c$ and $k_y$, showing band curvature decreasing until the band becomes flat band at $\mu_f = \mu_f^c$.

As we will see in the next section, the structure of the localized states in the flat band at energy $E_{fp}$ allows a percolation representation to be found for the interacting ground states. At any particular $f$ orbital site $\R$, only one localized state $|\psi_{\R, \sigma}\rangle$ is nonzero. This will be a valuable feature for studying the interacting system, since, using the localized states to construct many-body states, the Hubbard interaction at each $f$ orbital will give an energetic preference against doubly occupied states. 
The combination of Hund's coupling and Hubbard interactions on the $p$ orbital sites will then lead to spin exchange symmetry between overlapping localized states and thus to the percolation representation. 

\section{Ferromagnetic Percolation in Interacting Flat Bands}
\label{sec:perc}

Before discussing the main result, we first briefly review the percolation representation in the Hubbard model studied by Mielke and Tasaki \cite{Mielke1993a}. We then present and prove our results for the model discussed in Sec. \ref{sec:model}, where the presence of Hund's coupling plays an important role.

\subsection{Mielke-Tasaki Percolation in Flat Band Hubbard Models}

For a system where the lowest-energy single-particle band is flat, a percolation representation of ferromagnetism in the flat band was found in Ref. [\onlinecite{Mielke1993a}] in terms of a linearly independent set of $N_d$ states $\{\varphi_u(\r)\}$ spanning the space of the flat band. In this notation, $\r$ is  a lattice site and $u$ labels the state $\varphi_u$. The index set of all state labels $u$ is denoted $\Lambda_\varphi$. The percolation representation applies under two further conditions. First is \textit{quasilocality}, which requires each single-particle state to have a special site in its support where every other state vanishes, i.e., for each $u$ there is an $\r_u^*$ such that $\varphi_v(\r_u^*) \neq 0$ if and only if $v = u$. Second is that no more than two states can overlap at any site, i.e., for any site $\r$ there are at most two states $u$ and $v$ such that $\varphi_u(\r) \neq 0$ and $\varphi_v(\r) \neq 0$. Two states $u$ and $v$ are said to be \textit{directly connected} or \textit{overlapping} if there is such a site $\r$ where both states are nonzero.

Under these conditions, Ref. [\onlinecite{Mielke1993a}] found that if the number of electrons satisfies $N_e \leq N_d$, the ground states of the interacting system in the presence of a repulsive single-orbital Hubbard interaction can be written as a linear superposition of states formed from clusters of single-particle states with maximum total spin.

The ground state space being spanned by the cluster states yields a percolation representation where typical ground states are paramagnetic at low filling and ferromagnetic at high filling. If filling is low, a typical ground state consists of small, independent clusters that individually maximize spin but can have low total spin. If filling is high, there is typically one macroscopic cluster dominating the total spin. In particular, if $N_e = N_d$, the ground state space is spanned by fully spin-polarized states with trivial spin degeneracy. 
The above result from Ref. [\onlinecite{Mielke1993a}] thus allows for a filling-dependent ferromagnetic transition to be studied by treating the transition as a geometric percolation problem 
with additional spin weights due to the cluster $S_z$ degeneracy \cite{Maksymenko2012}.

This percolation representation was proven by constructing $N_e$-particle interacting states from single-particle flat band states and requiring the interacting states to have minimum Hubbard interaction energy, giving a ground state of the interacting system since the flat band minimizes the kinetic energy. In the proof of this result, the quasilocality condition is used to prevent double occupation of single-particle states, and the requirement that an overlap between single-particle states at a particular site involves only two states $u$ and $v$ yields the fact that the overall state is symmetric under swapping the spins of $u$ and $v$, indicating maximum total spin in the cluster. Violating either of these conditions can lead to clusters with lower total spin that still have zero interaction energy. 

For the model in Sec. \ref{sec:model}, the localized states shown in Fig. \ref{fig:model} satisfy quasilocality, as only one such state is nonzero at each $f$ orbital, but the conditions of the above theorem do not hold since three single-particle states have nonzero amplitude in the $p_y$ orbital component on each honeycomb lattice site $\r \in \Lambda_p$. Thus, just the interaction $H_U$ is insufficient to produce a percolation representation of maximum-spin clusters in our multiorbital model, and 
we will find that the combination of $H_U$ and $H_J$ allows for a percolation representation. We note as well that the $f$ orbitals allowing quasilocality to be satisfied in the basis of localized states $|\psi_{\R, \sigma}\rangle$ is necessary to find a percolation representation in terms of these localized states. 
Using features of the proof of our main result in Sec. \ref{sec:fmodelperc}, we demonstrate in Appendix \ref{appdx:lowerspin} an explicit example with a lower-than-maximum-spin cluster for the flat bands of the p-orbital Hamiltonian $H^p_K$ even when intraorbital Hubbard interactions and Hund's coupling are both considered.

\subsection{Hund's Coupling-Assisted Percolation Representation}
\label{sec:fmodelperc}

We will now see, by studying the model in Sec. \ref{sec:model}, that the presence of Hund's coupling can allow a percolation representation to be found for certain multiorbital systems even when there are overlaps between more than two localized states on one site.
The argument proceeds as follows. First, we perform a particle-hole transformation to study a model with a lowest-energy flat band. Next, we write an arbitrary ground state as a superposition of $N_e$-particle states in the localized state basis. Then, we require the ground state to be a zero-energy eigenstate of the the Hund's coupling as well as the intraorbital Hubbard interaction in the $f$, $p_x$, and $p_y$ orbitals. 
This will show that when the particle-hole-transformed flat band is at most half filled, the ground state space is spanned by states with maximum-spin clusters. Finally, we invert the particle-hole transformation to show that when the highest-energy flat band of the original system is at least half filled, the ground state space is also spanned by states with maximum-spin clusters.

Since the many-electron states used for the percolation representation will be constructed from the localized single-particle flat band states $|\psi_{\R, \sigma}\rangle$ in Eq. \eqref{eq:psi}, we begin by defining a creation operator $a^\dagger_{\R, \sigma}$ for the localized state $|\psi_{\R, \sigma}\rangle$ centered at site $\R \in\Lambda_f$,
\begin{equation}
\begin{aligned}
a^\dagger_{\R, \sigma}
&\equiv \sum_{\R\in\Lambda_f} \varphi_\R(\R, f) c^\dagger_\sigma(\R, f) + \sum_{\substack{\r \in\Lambda_p\\ p = p_x, p_y}} \varphi_\R(\r, p) c^\dagger_\sigma(\r, p)\\
&\equiv \vecg{\varphi}_{\R} \cdot \vec{c}^\dagger_{\sigma},
\label{eq:adagger}
\end{aligned}
\end{equation}
with $\vec{c}^\dagger_{\sigma}$ a compact notation for $f$-, $p_x$-, and $p_y$-orbital electron creation operators on the entire lattice $\Lambda$ as a  $(|\Lambda_f| + 2|\Lambda_p|)$-component vector. The wavefunction $\vecg{\varphi}_{\R}$ is compactly supported, taking nonzero values only for the $f$ orbital at $\R$ and the $p_x$ and $p_y$ orbitals at $\R + \u_i$ for $i = 1, 2, \dots, 6$, as shown in Fig. \ref{fig:state}. 
In this notation, a component of $\vec{c}^\dagger_\sigma$ will be labeled $c^\dagger_\sigma(\r, o_\r)$ where $\r$ is a site on the lattice and $o_\r$ is the orbital index at that site. Thus, if $\r = \R \in \Lambda_f$, $o_\r \in {\mathcal{O}}_\r = \{f\}$ and $c^\dagger_\sigma(\R, f) = f^\dagger_{\R, \sigma}$. Similarly, if $\r \in \Lambda_p$, $o_\r \in {\mathcal{O}}_\r = \{p_x, p_y\}$.
The dot product notation is a shorthand for summing over all sites and orbitals. The subscript $\R \in \Lambda_f$ in $\vecg{\varphi}_\R$ identifies the localized state centered at $\R$.

The maximum-spin cluster states that will span the ground state space for appropriate filling are defined by selecting a subset $A \subset \Lambda_f$ and placing one single-particle localized state $\vecg{\varphi}_\R$ at each $\R \in A$. The set $A$ can then be partitioned into $A = \bigsqcup_{k=1}^n C_k$, where $C_k$ are disjoint clusters, and two states labeled by $\R, \R' \in A$ belong to the same cluster if they are overlapping or connected by a path of overlapping states. Since total spin is maximized within each cluster, these states can be constructed from states $|\Phi_{A\uparrow}\rangle = \prod_{\R \in A} a^\dagger_{\R, \uparrow}|0\rangle$ or $|\Phi^{(h)}_{A\uparrow}\rangle = \prod_{\R \in A} a_{\R, \uparrow}|F \rangle$, depending on whether the flat band is the lowest- or highest-energy band. Here $|F\rangle = \prod_{\r \in \Lambda} \prod_{o_\r \in {\mathcal{O}_\r}} c_\uparrow^\dagger(\r, o_\r)c_\downarrow^\dagger(\r, o_\r)|0\rangle$ is the fully filled state and the orbital product is over ${\mathcal{O}_\r}$, the set of all orbitals $o_\r$ at site $\r$.

For the particle-hole transformed Hamiltonian, where the flat band is the lowest-energy band, the $S_z$ spin of each cluster can be lowered by a cluster spin-lowering operator $S^-_{C_k} = \sum_{(\r, o_\r) \in V_k} S^-_{\r, o_\r}$ with $S^-_{\r, o_\r} = c^\dagger_\downarrow(\r, o_\r) c_\uparrow(\r, o_\r)$ acting on $V_k = \{(\r, o_\r) | \varphi_\R(\r, o_\r) \neq 0 \mathrm{\ for \ any \ } \R \in C_k\}$, the set of orbitals where at least one state in the cluster is nonzero. For the original Hamiltonian, where the flat band is the highest-energy band, the analogous operator acting on holes is $S^{(h)-}_{C_k} = \sum_{(\r, o_\r)\in V_k} c_\downarrow(\r, o_\r)c^\dagger_\uparrow(\r, o_\r)$.

For the particle-hole transformed Hamiltonian, found by replacing creation and annihilation operators $f_{\R, \sigma} \leftrightarrow f^\dagger_{\R, \sigma}$, $p_{\r, \hat{x},\sigma}\leftrightarrow p^\dagger_{\r, \hat{x},\sigma}$, and $p_{\r, \hat{y},\sigma}\leftrightarrow p^\dagger_{\r, \hat{y},\sigma}$ in $H$, we find the following theorem.
\begin{theorem}
\label{thm:phtheorem}
Consider the particle-hole transformed Hamiltonian $H^{(ph)} = -H_K + H_U + H_J - \frac{J}{2}\hat{N}_p - H_\mu$ with $t_{fp}^2 > 3t_p t_f$ and $t_p, t_f, t_{fp}, U_p, U_f, J > 0$.  When $N_e \leq |\Lambda_f|$ and $\mu_f = \mu_f^{(ph), c} \equiv \mu_f^c + U_f/2 -U_p/2 + J/2$, with $\mu_f^c$ defined in Eq. \eqref{eq:mufc}, the ground state space of $H^{(ph)}$ is spanned by the states
\begin{equation}
    |\Phi^{(ph)}_{A, \{m_k\}}\rangle = \prod_{k = 1}^n (S_{C_k}^-)^{\frac{|C_k|}{2} - m_k} |\Phi_{A\uparrow}\rangle
    \label{eq:gsph}
\end{equation}
with $A\subset \Lambda_f$ and $|A| = N_e$.
\end{theorem}
One important detail is that the required $\mu_f$ in Theorem \ref{thm:phtheorem} is not the $\mu_f^c$ for which $H_K$ has a highest-energy flat band. Instead, there is a shift due to the additional chemical potential terms in the particle-hole-symmetric $H_U$ and the particle-hole-transformed Hund's coupling $H_J - \frac{J}{2}\hat{N}_p$. When $\mu_f = \mu_f^{(ph),c}$, the single-particle terms in $H^{(ph)}$ have a lowest-energy flat band spanned by the set of states $\{\vecg{\varphi}_\R\}_{\R \in \Lambda_f}$. We have discarded additional constant terms in $H^{(ph)}$, as they will not affect the spectrum or chemical potential condition.

For our original model, we find the following theorem.
\begin{theorem}
\label{thm:theorem}
Consider $H = H_K + H_U + H_J + H_\mu$ with $t_{fp}^2 > 3t_p t_f$ and $t_p, t_f, t_{fp}, U_p, U_f, J > 0$. When $N_e \geq 4|\Lambda_p| + |\Lambda_f|$ and $\mu_f = \mu_f^{(ph),c}$, defined in Theorem \ref{thm:phtheorem}, the ground state space of $H$ is spanned by 
\begin{equation}
    |\Phi_{A, \{m_k\}}\rangle = \prod_{k=1}^n (S^{(h)-}_{C_k})^{\frac{|C_k|}{2} -m_k} |\Phi^{(h)}_{A\uparrow}\rangle,
    \label{eq:gs}
\end{equation}
where $A\subset \Lambda_f$, $|A| = N_e$, and $A = \bigsqcup_{k=1}^n C_k$ where $C_k$ are disjoint clusters.
\end{theorem}

We now proceed with the proof of Theorem \ref{thm:phtheorem}, from which, our main result, Theorem \ref{thm:theorem} immediately follows by a particle-hole transformation. Importantly, the basis states $|\Phi_{A, \{m_k\}}\rangle$ feature maximum total spin within each cluster, since $|\Phi_{A, \{m_k\}}\rangle$ is a particle-hole transformation of $|\Phi^{(ph)}_{A, \{m_k\}}\rangle$ and total spin commutes with particle-hole transformations, as reviewed in Appendix \ref{appdx:spinph}.

\begin{proof} Following Ref. [\onlinecite{Mielke1993a}], we construct operators canonically conjugate to $a^\dagger_{\R, \sigma}$ by defining
\begin{equation}
    \begin{aligned}
    b_{\R, \sigma} &= \vecg{\kappa}_{\R}\cdot \vec{c}_{\sigma},\\
    \kappa_{\R}(\r, o_\r) &= \sum_{\R' \in \Lambda_f}(G^{-1})_{\R, \R'} \varphi_{\R'}(\r, o_\r),
    \end{aligned}
\end{equation}
with $G_{\R, \R'}=\vecg{\varphi}_{\R}\cdot\vecg{\varphi}_{\R'}$ the Gram matrix for the states $\vecg{\varphi}_{\R}$. Thus, $\vecg{\kappa}_\R\cdot\vecg{\varphi}_{\R'} = \delta_{\R, \R'}$ implies the canonical anticommutation relation $\{b_{\R', \sigma'}, a^\dagger_{\R, \sigma}\} = \delta_{\R, \R'} \delta_{\sigma, \sigma'}$. 
Note that since there is overlap between localized states centered on neighboring $f$ orbitals at $\R$ and $\R'$, $\{a_{\R',\sigma'}, a^\dagger_{\R,\sigma}\} \neq 0$, and $b_\R$ will serve as a more convenient annihilation operator. 
The states $\vecg{\kappa}_\R$ serve as an alternate basis for the flat band, and this Gram matrix procedure is essentially a method of constructing a dual basis where each element of the dual basis is orthogonal to all but one of the vectors in the original basis. This procedure is similar to the construction of reciprocal lattice vectors from direct lattice vectors.

Now the original electron operators $c_\sigma(\r, o_\r)$ can be expressed in terms of the operators $b_{\R}$ that annihilate the single-particle localized flat band states and operators $d_{\sigma}(\r, o_\r)$ orthogonal to the flat band. 
Multiplying the definition of $b_{\R}$ by $\varphi_\R(\r, o_\r)$ and summing over $\R$, $d_{\sigma}(\r, o_\r)$ can be defined to express the electron annihilation operators as
\begin{equation}
\begin{aligned}
    c_{\sigma}(\r, o_{\r}) = 
    \sum_{\R \in \Lambda_f} \varphi_{\R}(\r, o_\r) b_{\R, \sigma} + d_{\sigma}(\r, o_\r),
\end{aligned}
\end{equation}
where $d_\sigma(\r, o_\r) =\sum_{\r' \in \Lambda} \psi(\r, o_\r ; \r', o_{\r'})c_\sigma(\r', o'_{\r'})$ with 
$\psi(\r, o_\r ; \r', o'_{\r'})\equiv \delta_{\r, \r'}\delta_{o_\r, o'_{\r'}} - \sum_{\R\in\Lambda_f} \varphi_{\R}(\r, o_\r) \kappa_{\R}(\r', o'_{\r'})$ 
a projection out of the flat band spanned by $\vecg{\varphi_\R}$ and $\vecg{\kappa_\R}$. Thus, $\{d_\sigma(\r, o_\r), a^\dagger_{\R, \sigma}\} = \{d_\sigma(\r, o_\r), b^\dagger_{\R, \sigma}\} = 0$. Since the ground state will be expressed in terms of $a^\dagger_{\R, \sigma}$ operators, this construction allows the ground state condition in the presence of interactions to be analyzed using only states orthogonal or canonically conjugate to the single-particle localized flat band states.

When it is possible to construct an $N_e$-electron state from single-particle flat band states that simultaneously minimizes the interactions, such a state will be a ground state, and the ground state space will be spanned by the collection of these states. In particular, the $n_\uparrow n_\downarrow$ interaction $H_U'$ and $H_J$ are positive semidefinite, so when there are states satisfying $H_U'|\Phi\rangle = H_J|\Phi\rangle = 0$ with energy $-N_e E_{fp}$, these states will be ground states of $H^{(ph)}$ and can be written
\begin{equation}
    |\Phi^{(ph)}\rangle = \sum_{A_\uparrow, A_\downarrow\subset \Lambda_f} f(A_\uparrow, A_\downarrow)\prod_{\R\in A_\uparrow\cup A_\downarrow}\prod_{\sigma_\R} a^\dagger_{\R, \sigma_{\R}} |0\rangle,
\end{equation}
with constraints on the coefficients $f(A_\uparrow, A_\downarrow)$ to be determined by the zero-interaction-energy conditions. Here $A_\uparrow$ and $A_\downarrow$ are subsets of $\Lambda_f$ and $\sigma_\R = \uparrow$ or $\downarrow$ if $\R \in A_\uparrow$ or $\R \in A_\downarrow$. If $\R \in A_\uparrow \cap A_\downarrow$, the product over $\sigma_\R$ includes both, with the spin-up operator to the left. The sum over $A_\uparrow, A_\downarrow$ is a sum over all possible such subsets satisfying $|A_\uparrow| + |A_\downarrow| = N_e$. We will see that $|\Phi^{(ph)}\rangle = 0$ if $N_e > |\Lambda_f|$.

Since $H_U'$ and $H_J$ are themselves sums of positive semidefinite operators at each site, we first consider the Hubbard interaction on an $f$ orbital at $\R$, which gives the condition
\begin{equation}
    \begin{aligned}
   0 &= c_{\uparrow}(\R, f)c_{\downarrow}(\R, f) |\Phi^{(ph)}\rangle \\
   &= \left(d_\uparrow(\R, f) + \sum_{\R'\in\Lambda_f} \varphi_{\R'}(\R, f)b_{\R',\uparrow} \right) \\
    &\times \left(d_\downarrow(\R, f) +\sum_{\R'' \in \Lambda_f} \varphi_{\R''}(\R, f)b_{\R'', \downarrow}\right)|\Phi^{(ph)}\rangle \\
   &\implies b_{\R, \uparrow}b_{\R, \downarrow}|\Phi^{(ph)}\rangle = 0,
    \end{aligned}
    \label{eq:fcond}
\end{equation}
since $d_\sigma(\r, o_\r)$ anticommutes with $b_{\R, \sigma}$ and with the $a^\dagger_{\R, \sigma}$ operators in $|\Phi\rangle$, and $\varphi_{\R'}(\R, f) \neq 0$ only when $\R' = \R$. Thus, since this condition holds for any $\R \in \Lambda_f$, the state $|\Phi^{(ph)}\rangle$ must satisfy $f(A_\uparrow, A_\downarrow) = 0$ if $A_\uparrow \cap A_\downarrow \neq \emptyset$, or, in other words, there must be no double occupancy of localized single-particle states. If $N_e > |\Lambda_f|$, this condition can only be met if $|\Phi^{(ph)}\rangle = 0$, meaning the ground state cannot be expressed solely in terms of $N_e$ flat band states and must have energy higher than $-N_e E_{fp}$.

Next, we examine the Hubbard interaction in the $p_x$ and $p_y$ orbitals. Consider a honeycomb cell labeled by $\R_0 \in \Lambda_f$ and examine its rightmost vertex, $\r_0 = \R_0 + \u_1$. Two additional honeycomb cells, centered at $\R_1 = \R_0 + \w_1$ and $\R_6 = \R_0 + \w_6$, share vertex $\r_0$. The corresponding localized single-particle states $\vecg{\varphi}_{\R_i}$ have nonzero component in the $p_y$ orbital at $\r_0$ for all three of $\R_0$, $\R_1$, and $\R_6$, but only $\R_1$ and $\R_6$ have a nonzero $p_x$ component at $\r_0$. Excluding the normalization factor $t_f/\sqrt{6(t_f^2 + t_{fp}^2)}$ on the $p$-orbital components, the nonzero $p_x$ components are, $\varphi_{\R_1}(\r_0, p_x) = -\frac{\sqrt{3}}{2}$ and $\varphi_{\R_6}(\r_0, p_x) = \frac{\sqrt{3}}{2}$ while the nonzero $p_y$ components are $\varphi_{\R_0}(\r_0, p_y) = 1$, $\varphi_{\R_1}(\r_0, p_y) = -\frac{1}{2}$, and $\varphi_{\R_6}(\r_0, p_y) = -\frac{1}{2}$. No other localized states $\vecg{\varphi}_\R$ are nonzero at $\r_0$.

The zero-interaction-energy condition for the $p_x$-orbital Hubbard interaction at site $\r_0$ is
\begin{equation}
\begin{aligned}
   0 &= c_\uparrow(\r_0, p_x)c_\downarrow(\r_0, p_x) |\Phi^{(ph)}\rangle\\
     &= \left(\sum_{i = 1, 6} \varphi_{\R_i}(\r_0, p_x) b_{\R_i, \uparrow}\right) \\
     &\times \left(\sum_{i = 1, 6} \varphi_{\R_i}(\r_0, p_x) b_{\R_i, \downarrow}\right)|\Phi^{(ph)}\rangle \\
     &\implies (b_{\R_1, \uparrow} b_{\R_6, \downarrow}- b_{\R_1, \downarrow} b_{\R_6, \uparrow}  ) |\Phi^{(ph)}\rangle = 0,
\end{aligned}
\label{eq:pxcond}
\end{equation}
where the last line follows from the no-double-occupancy condition Eq. \eqref{eq:fcond} and the fact that $\varphi_{\R_0}(\r_0, p_x) = 0$. The condition in Eq. \eqref{eq:pxcond} essentially projects out states involving a spin-singlet component between the states centered at $\R_1$ and $\R_6$. Explicitly, this condition gives that for any configuration where $A_\uparrow = B_\uparrow\sqcup\{\R_1\}$ and $A_\downarrow = B_\downarrow\sqcup\{\R_6\}$ with $\R_1, \R_6 \notin B_\uparrow,B_\downarrow$, the coefficients are symmetric under exchange of spins, $f(B_\uparrow\sqcup\{\R_1\}, B_\downarrow\sqcup\{\R_6\}) = f(B_\uparrow\sqcup\{\R_6\}, B_\downarrow\sqcup\{\R_1\})$. Since the choice of $\R_0$ is arbitrary, this gives the general condition that the state must have spin exchange symmetry between localized single-particle states that are nearest neighbors in the $y$ direction, states $\R$ and $\R' = \R + \w_2$.  Eqs. \eqref{eq:fcond} and \eqref{eq:pxcond} are equivalent to the conditions resulting from quasilocality and the requirement that no more than two single-particle states overlap at any site in Ref. [\onlinecite{Mielke1993a}].

The zero-interaction-energy condition for the $p_y$ orbitals involves three overlapping states at any site. For the $p_y$ orbital at site $\r_0$, using Eqs. $\eqref{eq:fcond}$ and  $\eqref{eq:pxcond}$, 
\begin{equation}
\begin{aligned}
   0 &= c_\uparrow(\r_0, p_y)c_\downarrow(\r_0, p_y) |\Phi^{(ph)}\rangle\\
     &= \varphi_{\R_6}(\r_0, p_y)\varphi_{\R_0}(\r_0, p_y)(b_{\R_6, \uparrow} b_{\R_0, \downarrow} - b_{\R_6, \downarrow} b_{\R_0, \uparrow})|\Phi^{(ph)}\rangle
     \\&+\varphi_{\R_1}(\r_0, p_y)\varphi_{\R_0}(\r_0, p_y)(b_{\R_1, \uparrow} b_{\R_0, \downarrow} - b_{\R_1, \downarrow} b_{\R_0, \uparrow})|\Phi^{(ph)}\rangle \\
     &\implies (b_{\R_6, \uparrow} b_{\R_0, \downarrow} - b_{\R_6, \downarrow} b_{\R_0, \uparrow})|\Phi^{(ph)}\rangle
     \\&\hspace{0.8cm}+(b_{\R_1, \uparrow} b_{\R_0, \downarrow} - b_{\R_1, \downarrow} b_{\R_0, \uparrow})|\Phi^{(ph)}\rangle
     = 0.
\end{aligned}
\label{eq:pycond}
\end{equation}
While this condition is satisfied for states that maximize spin (have spin exchange symmetry) between single-particle states at $\R_6$ and $\R_0$ as well as between those at $\R_1$ and $\R_0$, it is not the case that every state $|\Phi\rangle$ satisfying Eq. \eqref{eq:pycond} must have such spin exchange symmetry. Thus, the percolation representation is not strictly valid when the only interaction is $H_U'$.

The final condition we consider is $H_J|\Phi\rangle = 0$. This condition can be written using only annihilation operators in $H_J$ by writing $H_J = \frac{J}{2}\sum_{\r \in \Lambda_p} n_{\r, S=0}$, where $n_{\r, S=0} = (\pydown^\dagger\pxup^\dagger  - \pyup^\dagger\pxdown^\dagger )(\pxup \pydown - \pxdown \pyup) \equiv c^\dagger_{\r, S=0} c_{\r, S=0}$ is an operator that counts whether there is a spin singlet between the $p_x$ and $p_y$ orbitals at site $\r$. The derivation of this operator identity is shown in Appendix \ref{appdx:Hund}. In this form, $H_J$ is clearly a sum of positive semidefinite operators, and $H_J|\Phi\rangle = 0$ if and only if $c_{\r, S=0}|\Phi\rangle = 0$ for every $\r \in \Lambda_p$.

In terms of localized state operators, the zero-interaction-energy condition for Hund's coupling at site $\r_0$ is
\begin{equation}
    \begin{aligned}
    0 &= \left[c_\uparrow(\r_0, p_x)c_\downarrow(\r_0,p_y) -c_\downarrow(\r_0, p_x)c_\uparrow(\r_0,p_y)\right]|\Phi^{(ph)}\rangle \\
    &= \varphi_{\R_6}(\r_0,p_x) \varphi_{\R_0}(\r_0, p_y) (b_{\R_6, \uparrow} b_{\R_0, \downarrow} - b_{\R_6, \downarrow} b_{\R_0, \uparrow})|\Phi^{(ph)}\rangle \\
    &+\varphi_{\R_1}(\r_0, p_x) \varphi_{\R_0}(\r_0, p_y)(b_{\R_1, \uparrow} b_{\R_0, \downarrow} - b_{\R_1, \downarrow} b_{\R_0, \uparrow})|\Phi^{(ph)}\rangle \\
    &\implies (b_{\R_6, \uparrow} b_{\R_0, \downarrow} - b_{\R_6, \downarrow} b_{\R_0, \uparrow})|\Phi^{(ph)}\rangle
     \\&\hspace{0.8cm}-(b_{\R_1, \uparrow} b_{\R_0, \downarrow} - b_{\R_1, \downarrow} b_{\R_0, \uparrow})|\Phi^{(ph)}\rangle = 0,
    \label{eq:hjcond}
    \end{aligned}
\end{equation}
using the no-double-occupancy condition from Eq. \eqref{eq:fcond} and the spin triplet condition between $\R_1$ and $\R_6$ from Eq. \eqref{eq:pxcond}. Taking the sum and difference of the conditions in Eqs. \eqref{eq:pycond} and \eqref{eq:hjcond} gives spin triplet conditions between $\R_0$ and $\R_1$ and between $\R_0$ and $\R_6$. 
In other words, the spin degree of freedom must be fully symmetrized among any overlapping states at $\r_0$. This argument holds for any choice of $\R_0$, meaning it applies at any site in the same honeycomb sublattice as $\r_0$. 
In fact, as can be seen by considering the left-most site on the $\R_0$-centered honeycomb, $\r_0' = \R_0 + \u_4$, these spin symmetrization conditions apply both sublattices of the $p$-orbital honeycomb lattice $\Lambda_p$. If spin is symmetrized between any two overlapping localized states, a cluster of localized states will be fully spin symmetrized, since any two localized states in a cluster can be connected by a path of overlapping localized states each adjacent pair of which must have symmetrized spin. Thus, clusters of localized states have maximum total spin $S_{C_k, tot} = \frac{|C_k|}{2}$ and ground states for $N_e \leq |\Lambda_f|$ can be written in the form of Eq. \eqref{eq:gsph}.
\end{proof}

Theorem \ref{thm:theorem} follows immediately by particle-hole transformation. In particular, since Theorem \ref{thm:phtheorem} requires the lowest-energy flat band to be \textit{at most} half filled, Theorem \ref{thm:theorem} requires the highest-energy flat band to be \textit{at least} half filled. The clusters in Eq. \eqref{eq:gs} are then connected sets of holes, or singly-occupied localized states, surrounded by a doubly-occupied background. 

In order to interpret Theorem \ref{thm:theorem} as a percolation representation, note that when the highest-energy flat band is exactly half filled, $|N_e| = 4|\Lambda_p| +|\Lambda_f|$, there is a single cluster spanning the system, and all ground states have total spin $\frac{|\Lambda_f|}{2}$. When the system is close to fully filled, clusters are small and it is easy to find combinations of basis states in Eq. \eqref{eq:gs} with low total spin. As the highest-energy flat band approaches half filling, the ground state space becomes dominated by states with a large cluster spanning the system and carrying large spin. There is thus a paramagnetic-ferromagnetic transition as the system approaches half filling of the top band from above in the sense that sufficiently close to half filling, the ground state space is dominated by states with macroscopic spin.

\subsection{Monte Carlo Simulation of the Correlated Percolation Transition}

With the result in Theorem \ref{thm:theorem}, we can now study the ferromagnetic transition in our multiorbital model through Monte Carlo simulations for correlated percolation.
The ground state basis in Theorem \ref{thm:theorem} is first reorganized into a purely geometric percolation representation 
by averaging over the cluster configurations and spins \cite{Mielke1993a, Maksymenko2012}. 
At a fixed filling, whether typical states in the ground state space are ferromagnetic can be determined by considering the total spin $S^2$ averaged over the ground state space. Fixing the cluster configuration $A$ and averaging over the $S^z$ spins $\{m_k\}$ of the clusters gives that the averaged total spin of the cluster configuration, $S_A$, depends only on the size of each cluster $C_k \subset A$ \cite{Mielke1993a, Maksymenko2012},
\begin{equation}
\begin{aligned}
    S_A^2 &\equiv \frac{1}{W(A)} \sum_{\{m_k\}} \langle \Phi_{A, \{m_k\}}| S^2 | \Phi_{A, \{m_k\}}\rangle \\
    &= \sum_{k=1}^n \frac{|C_k|}{2}\left(\frac{|C_k|}{2} + 1\right).
\label{eq:spin}
\end{aligned}
\end{equation}
This result can be interpreted as a geometric correlated percolation representation for the averaged ground state spin. For our model, the percolation problem is defined on the triangular lattice $\Lambda_f$ where a geometric configuration is specified by the set of filled sites $A\subset \Lambda_f$ and each configuration has a weighting factor $W(A) = \sum_{\{m_k\}} 1 = \prod_{k=1}^n (|C_k| + 1)$. 
Since Theorem \ref{thm:theorem} applies when the highest band in the multiorbital system is at least half filled, filled sites in the percolation representation correspond to holes, or singly occupied localized states, in the multiorbital system. Similarly, empty sites in the percolation representation correspond to doubly occupied localized states in the multiorbital system. In the following discussion, we will refer directly to filled and empty sites in the percolation representation, with filling density $p$ corresponding to the hole density per unit cell of the original model.

At fixed filling $p$, which defines the canonical ensemble, the spin averaged over all cluster configurations $A$ in the ground state space is
\begin{equation}
\begin{aligned}
    \langle S^2 \rangle &= \frac{\sum_A W(A) S_A^2}{Z}, \\
    W(A) &= \prod_{k=1}^n \left(|C_k|+1\right),\\
    Z &= \sum_A W(A),
\label{eq:weights}
\end{aligned}
\end{equation}
with the sum over subsets $A \subset \Lambda_f$ with $|A| = N = p|\Lambda_f|$, and the weights $W(A)$ account for cluster $S^z$ degeneracy. When examining the ensemble averaged spin, it will be useful to consider the spin fraction $\overline{s^2} \equiv \langle S^2 \rangle / S^2_{\text{max}}$, where $S^2_{\text{max}} = N/2 (N/2 + 1)$ corresponds to a configuration with all filled sites in a single cluster.

It will also be useful to consider the grand canonical ensemble at fixed fugacity $z \equiv e^{\tilde{\mu}}$, which requires the grand canonical weights
\begin{equation}
    W_{\text{GC}}(A) = \prod_{k=1}^n z^{|C_k|} (|C_k|+1)
\label{eq:gcweights}
\end{equation}
and the corresponding ensemble average involving a sum over all subsets $A \subset \Lambda_f$ without fixing $|A|$.
The fugacity $z$ and the corresponding $\tilde{\mu}$ used in the grand canonical simulations couple directly to the percolation representation, or to the holes in the flat band, and are thus not related to the physical chemical potentials $\mu_p$ and $\mu_f$.

For canonical ensemble simulations, filling is fixed, and the update step consists of an attempted swap of a randomly selected filled site with a randomly selected empty site. For grand canonical ensemble simulations at fixed $z$, an update step consists of randomly selecting a site and attempting to fill it if empty or empty it if filled. A proposed update from configuration $A$ to $A'$ is then accepted with probability $\text{min}\{1,W(A')/W(A)\}$ or $\text{min}\{1,W_{\text{GC}}(A')/W_{\text{GC}}(A)\}$, depending on the ensemble. In both cases, simulations are done on the triangular lattice with periodic boundary conditions. Further details about the algorithm used for correlated percolation can be found in Appendix \ref{appdx:alg}. 

\begin{figure}[tbp]
\subfigure[]{\epsfig{file=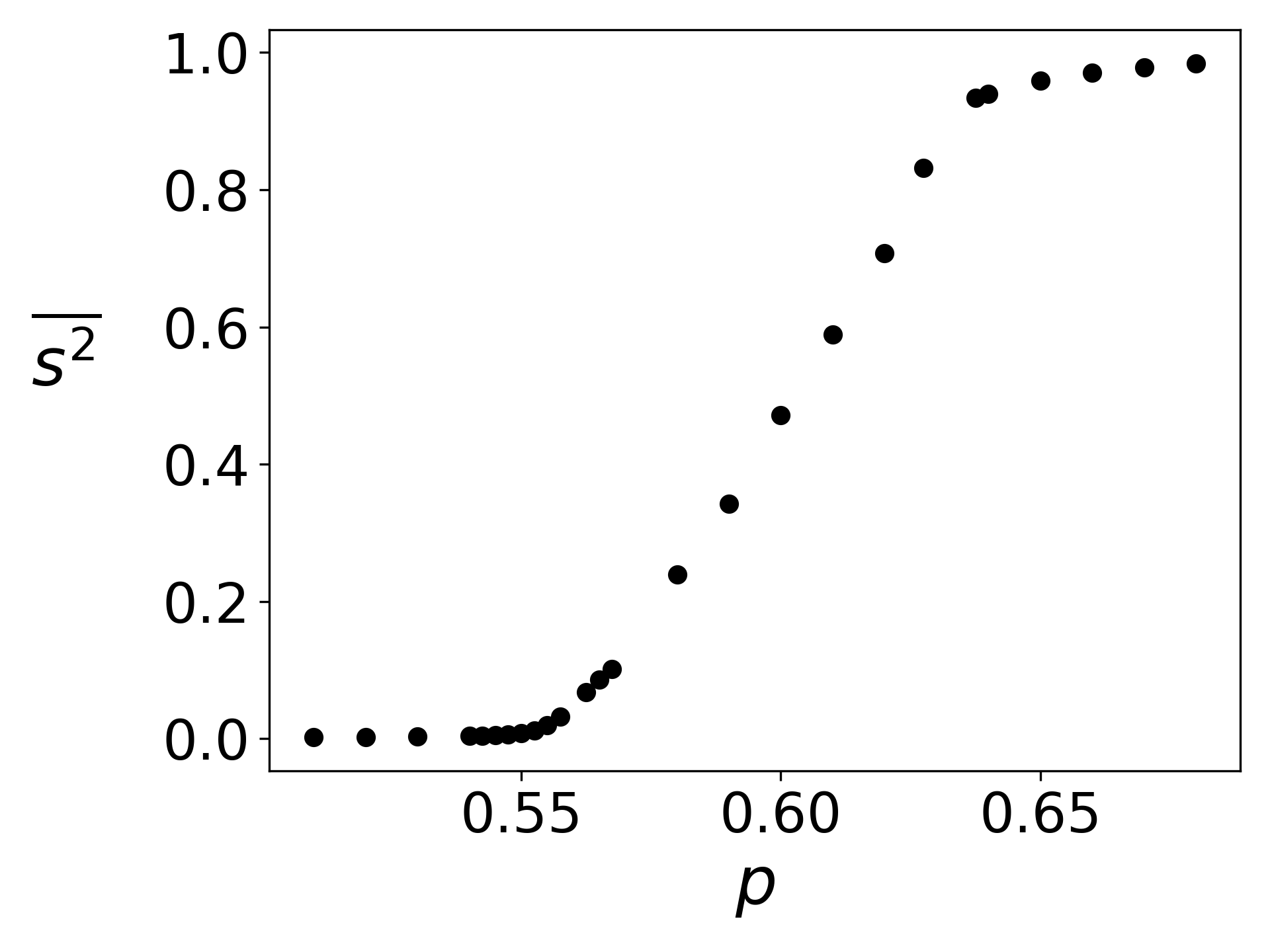, width=0.49\linewidth}}
\subfigure[]{\epsfig{file=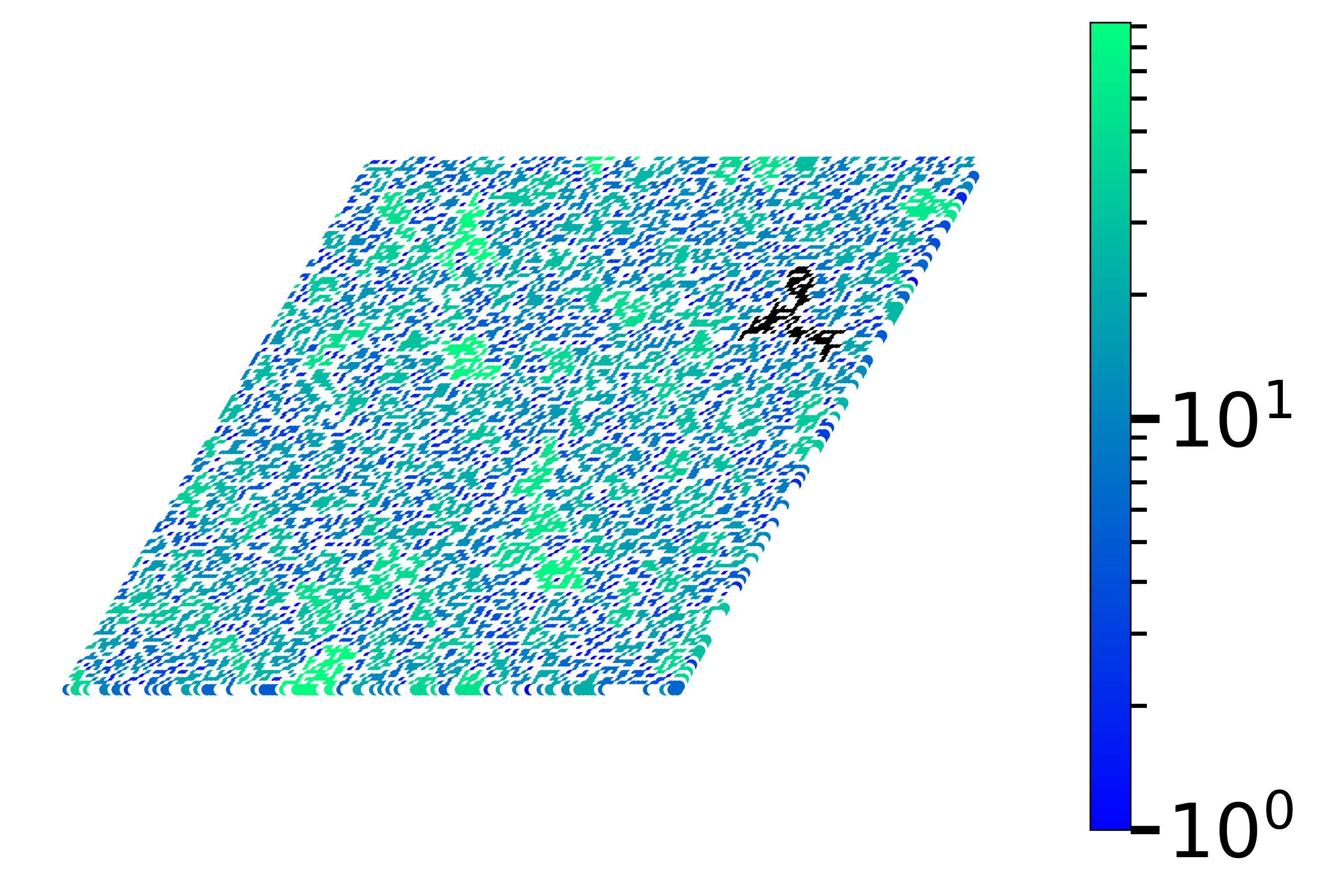, width=0.49\linewidth}}
\subfigure[]{\epsfig{file=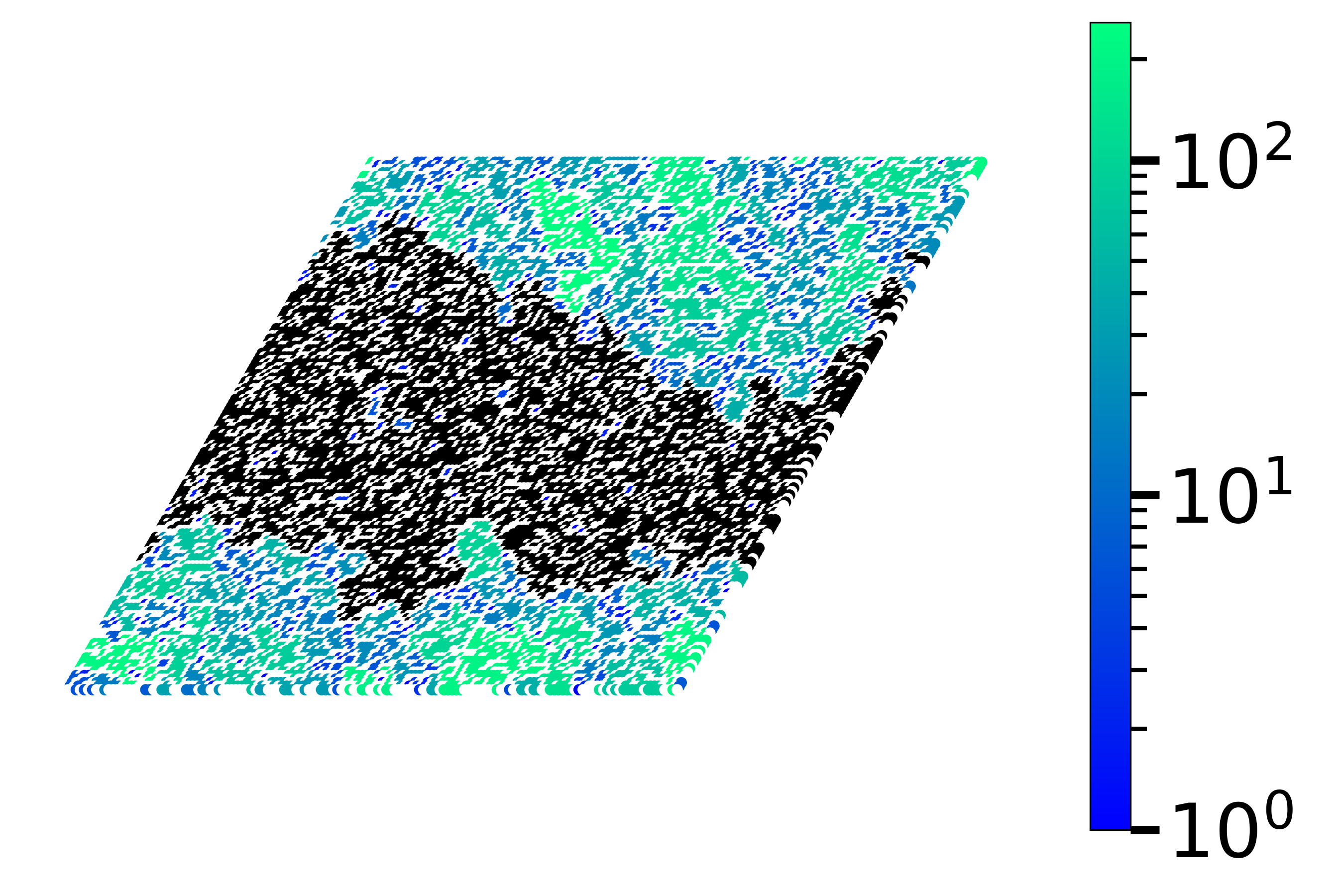, width=0.49\linewidth}}
\subfigure[]{\epsfig{file=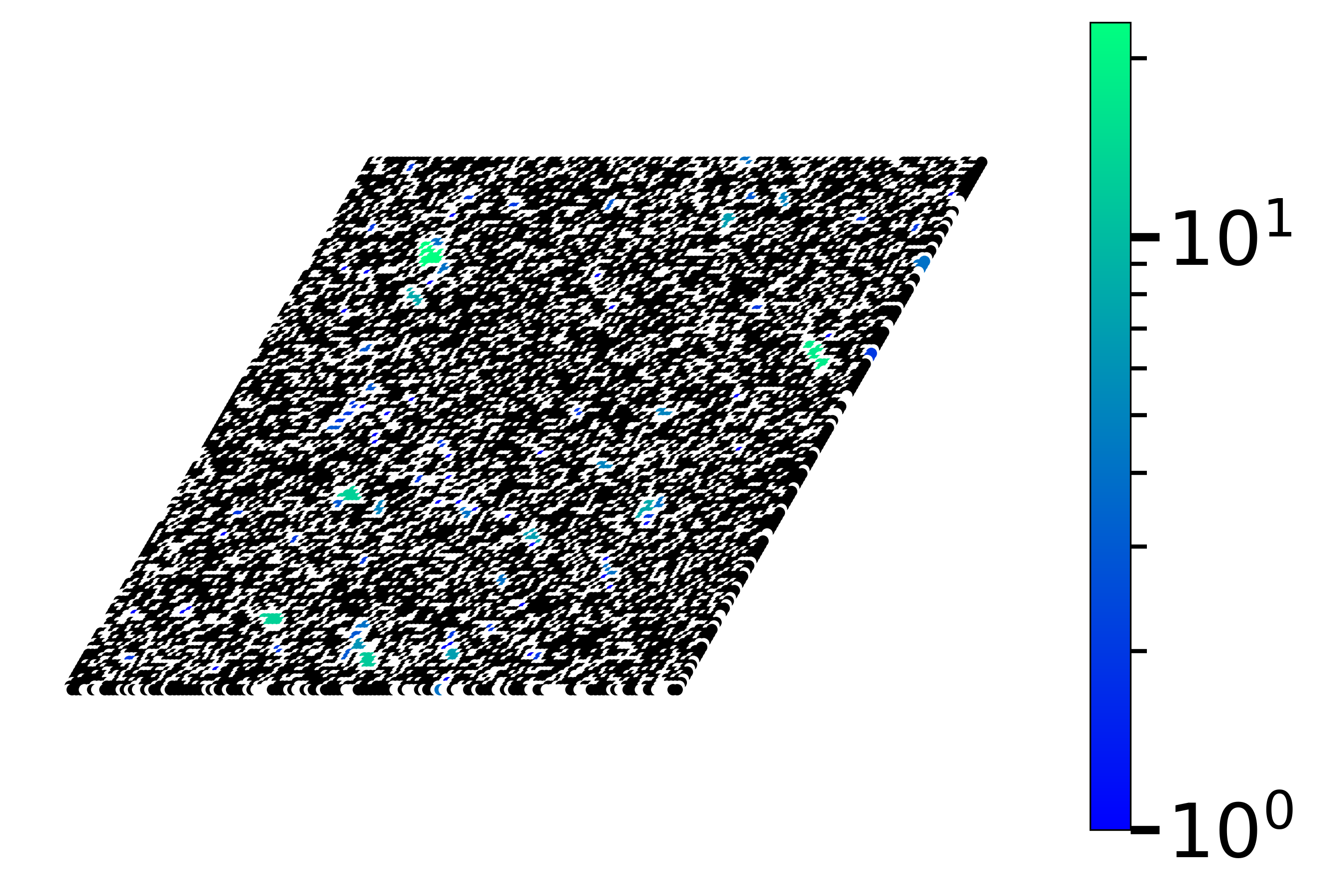, width=0.49\linewidth}}
\caption{(a) Averaged spin fraction $\overline{s^2}$ in the canonical ensemble at different fillings $p$ on a triangular lattice of size $|\Lambda_f| = 230 \times 230$. Typical configurations are shown on a $150\times 150$ system for (b) $p = 0.45$, (c) $p = 0.58$, and (d) $p = 0.65$. Empty sites are white, and the largest cluster is black. The remaining clusters are colored on a blue-green scale corresponding to cluster size.}
\label{fig:canonical}
\end{figure}

Results for the canonical ensemble in a range of fillings near the transition are shown in Fig. \ref{fig:canonical}(a). The spin fraction $\overline{s^2}$ is near zero, and in fact scales to zero with system size, when $p < p_{1} \approx 0.55$, and the system is in a paramagnetic phase with typical configurations having small clusters, as shown for $p=0.45$ in Fig. \ref{fig:canonical}(b). For $p_{1} < p < p_{2} \approx 0.63$, the spin fraction grows quickly with $p$ and typical configurations are phase separated, with a region with small clusters separated from a region with a macroscopic cluster shown for $p = 0.58$ in Fig. \ref{fig:canonical}(c). This phase separation behavior, which does not appear in the standard site percolation problem, has been previously observed for correlated percolation on the square lattice in Ref. \cite{Maksymenko2012}, where it was understood by interpreting the weights as an effective repulsive interaction.
For $p > p_{2}$, the largest cluster spreads uniformly throughout the lattice, as shown for $p = 0.65$ in Fig. \ref{fig:canonical}(d), and the system is ferromagnetic. As $p \to 1$, the system becomes fully spin polarized. Due to the effective repulsive interaction of the weights, the fillings $p_1$ and $p_2$ are both larger than the critical filling $p_c = 0.5$ for standard site percolation on the triangular lattice \cite{Grimmett1999}. 
Since the filling $p$ in the percolation representation corresponds to hole density in the original multiorbital model, the corresponding electron densities per unit cell for the paramagnetic phase are $n_e > 9.45$, for phase separation are $9.45 > n_e > 9.37$, and for the ferromagnetic phase are $9.37 > n_e \geq 9$ with fully saturated ferromagnetism at flat band half filling $n_e = 9$.

\begin{figure}[tbp]
\epsfig{file=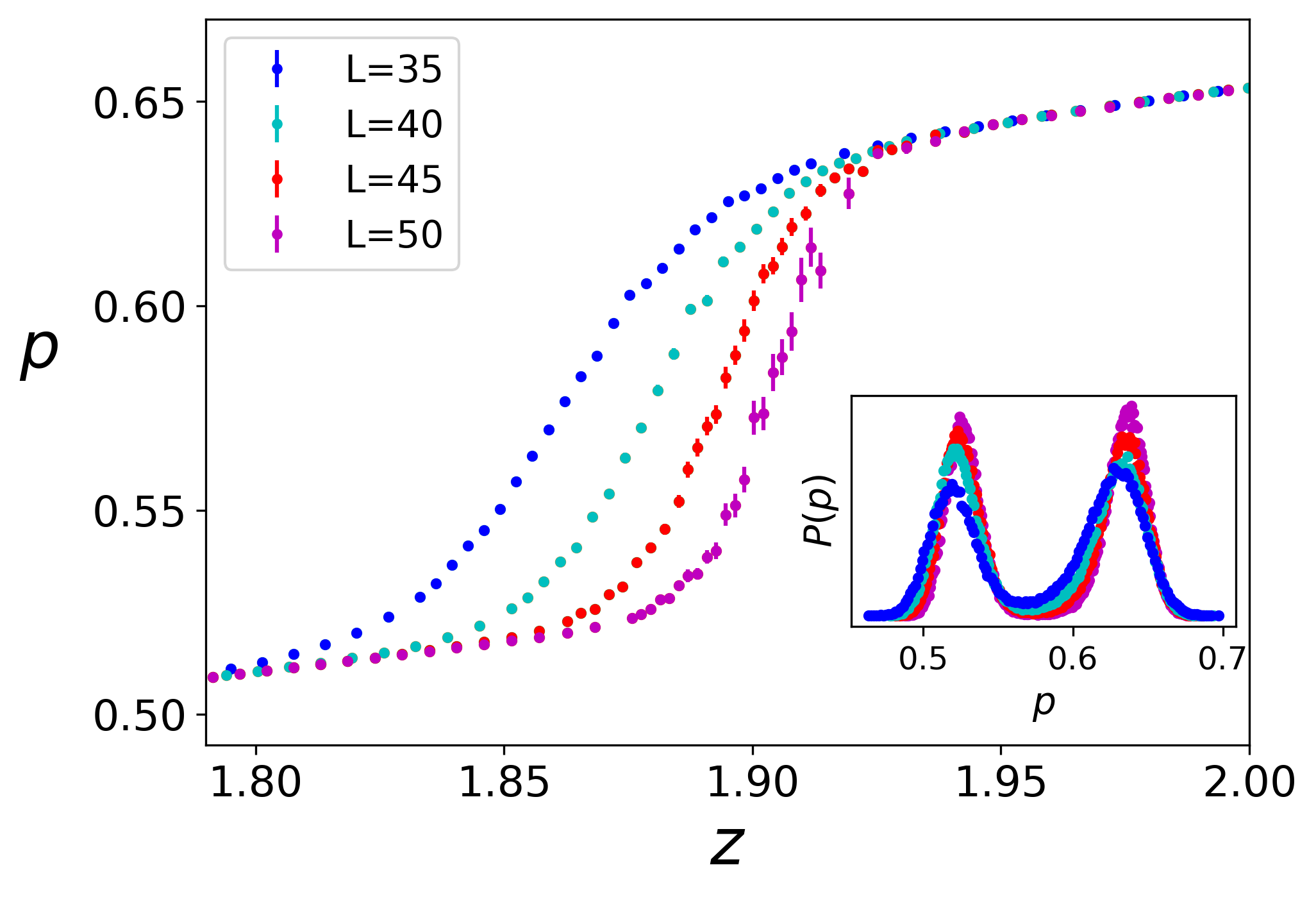, width=\linewidth}
\caption{Filling $p$ as a function of fugacity $z$ in the grand canonical ensemble with different system sizes $|\Lambda_f| = L\times L$ shown in different colors. Inset: Histograms of filling at the finite size estimates of $z_c$ for each system size. The histograms use bins of width $1/L^2$ and are rescaled by the bin width to give an approximation of the probability density $P(p)$ that a configuration is at filling $p$ for the particular $z_c$. Any empty bins are excluded.}
\label{fig:gc}
\end{figure}

For grand canonical ensemble simulations, filling $p$ as a function of flat band fugacity $z$ is shown in Fig. \ref{fig:gc} for triangular lattices with linear dimension $L \equiv \sqrt{|\Lambda_f|}$. As the system size increases, the transition sharpens, suggesting a discontinuity in $p(z)$ at a critical $z_c$ in the thermodynamic limit. For finite sizes, the estimate of $z_c$ is given by the point at which the histogram of the Monte Carlo configurations features two well defined peaks with approximately equal height, as shown in the inset of Fig. \ref{fig:gc}.  For $L=50$, jumps between
fillings $p^{gc}_{1} = 0.53(1)$ and $p^{gc}_{2} = 0.64(1)$ occur at the estimated $z_c \approx 1.91$.

\section{Conclusions}

We have established an exact result useful for studying ferromagnetism in an interacting multiorbital flat band system. 
We constructed a multiorbital Hubbard model on a two-layer lattice consisting of a honeycomb lattice layer of $p_x$ and $p_y$ orbitals and a triangular lattice layer of $f_{y(3x^2-y^2)}$ orbitals aligned with the centers of the honeycomb plaquettes. 
For an appropriate chemical potential difference between the two layers, the model exhibits two flat bands due to destructive interference. 
The presence of Hund's coupling between degenerate $p$ orbitals, in addition to repulsive intraorbital Hubbard interactions, allows the highest-energy flat band to admit a provable percolation representation for the degenerate many-body ground states.
Ground states correspond to configurations of spin-polarized clusters of localized electron states, leading to a percolation representation for the average ground state spin.
As the flat band filling varies from fully filled to half filled, the many-body ground states transition from a paramagnetic to a ferromagnetic phase, as shown by Monte Carlo simulations for correlated percolation.

\section*{Acknowledgments}
We thank Tyrel McQueen for helpful discussion about potential material realizations. 
Simulations were performed using the Maryland Advanced Research Computing Center (MARCC) Blue Crab cluster. 
We acknowledge the use of the pymbar package \cite{Shirts2008, Chodera2016} during data processing. 
This work is supported by the NSF CAREER grant DMR-1848349 and in part by the Alfred P. Sloan Research Fellowships under grant FG-2018-10971.

\appendix
\section{Lower-spin Clusters in $p$-orbital System}
\label{appdx:lowerspin}
In this appendix, we consider just the $p_x$- and $p_y$-orbital Hamiltonian $H = H_K^p + H_U^p + H_J$, including the $p$-orbital kinetic Hamiltonian in Eq. \eqref{eq:hp}, the intraorbital Hubbard interactions in the $p$ orbitals $H_U^p=U_p \sum_{\r\in\Lambda_p} \sum_{p = p_x, p_y} \left(n_{\r, p, \uparrow} -\frac{1}{2}\right)\left(n_{\r, p, \downarrow}-\frac{1}{2}\right)$, as well as Hund's coupling between $p$ orbitals in Eq. \eqref{eq:hj}. By use of a simple example, we show that the loop state basis $|\psi_{\R, \sigma}^{-(p)}\rangle$ does not admit a simple percolation representation, as there are states where clusters do not maximize total spin. For consistency, we use the notation from Sec. \ref{sec:fmodelperc} of the main text with the modification that there are no $f$ orbital sites and $\Lambda_f$ is simply the set of honeycomb plaquette labels. Thus, $\varphi_\R(\r, o_\r)$ describes the component at orbital $o_\r = p_x, p_y$ of site $\r$ of a loop state on the plaquette $\R$. All operators $a_{\R, \sigma}$ and $b_{\R, \sigma}$ are defined similarly to those in Sec. \ref{sec:fmodelperc} using the wavefunctions $\varphi_\R(\r, o_\r)$ defined only at sites $\r\in\Lambda_p$.

Consider an arbitrary plaquette centered at $\R_0$ together with the six surrounding plaquettes centered at $\R_i = \R_0 + \w_i$ with $i = 1, \dots, 6$. The nonzero components at site $\r_1 = \R_0 + \u_1$ are $\varphi_{\R_1}(\r_1, p_x) = -\frac{\sqrt{3}}{2}$ and $\varphi_{\R_6}(\r_1, p_x) = \frac{\sqrt{3}}{2}$ while the nonzero $p_y$ components are $\varphi_{\R_0}(\r_1, p_y) = 1$, $\varphi_{\R_1}(\r_1, p_y) = -\frac{1}{2}$, and $\varphi_{\R_6}(\r_1, p_y) = -\frac{1}{2}$, discarding the $1/\sqrt{6}$ normalization factor. 

\begin{figure}[tbp]
\epsfig{file=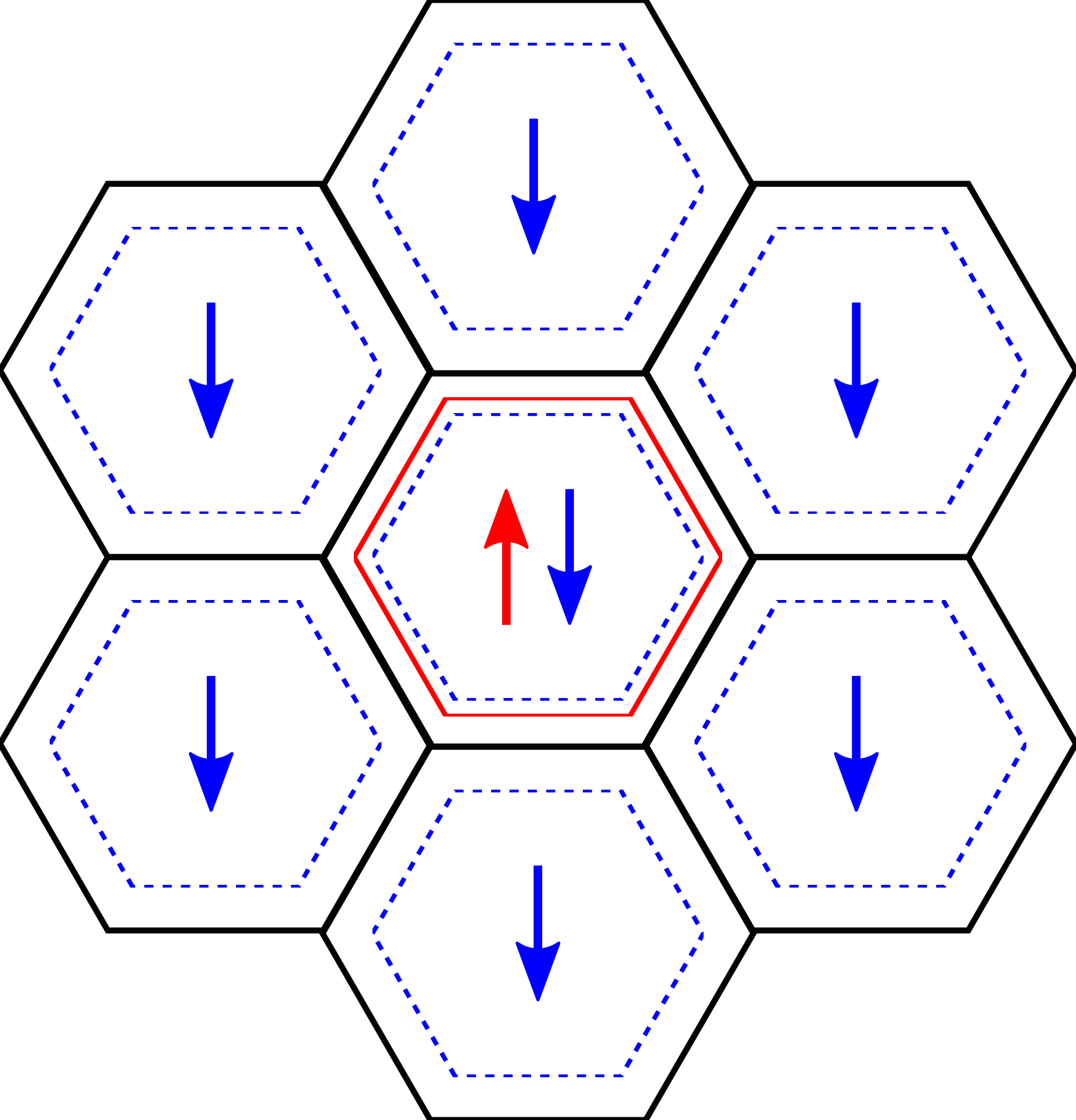, width=0.48\linewidth}
\caption{Two-electron state $|\Phi'_{\R_0}\rangle$, which consists of a central filled spin-up loop surrounded by a superposition of spin-down loops on the central and six surrounding plaquettes. This state does not maximize spin despite avoiding the interaction energy.}
\label{fig:spinexample}
\end{figure}

The equivalents of the zero-interaction-energy conditions in Eq. \eqref{eq:pxcond}, \eqref{eq:pycond}, and \eqref{eq:hjcond} in this case are
\begin{equation}
    \begin{aligned}
        0 = &\big[b_{\R_1\uparrow}b_{\R_1\downarrow} + b_{\R_6\uparrow}b_{\R_6\downarrow} 
        - (b_{\R_1\uparrow}b_{\R_6\downarrow} - b_{\R_1\downarrow}b_{\R_6\uparrow})\big]|\Phi\rangle \\
        0 = &\big[4b_{\R_0\uparrow} b_{\R_0\downarrow} + b_{\R_1\uparrow}b_{\R_1\downarrow} + b_{\R_6\uparrow}b_{\R_6\downarrow}\\
        &+(b_{\R_1\uparrow}b_{\R_6\downarrow} - b_{\R_1\downarrow}b_{\R_6\uparrow}) -2(b_{\R_0\uparrow} b_{\R_6\downarrow} - b_{\R_0\downarrow}b_{\R_6\uparrow}) \\
        &-2(b_{\R_0\uparrow}b_{\R_1\downarrow} - b_{\R_0\downarrow}b_{\R_1\uparrow})\big]|\Phi\rangle \\
        0 = &\big[b_{\R_1\uparrow}b_{\R_1\downarrow} - b_{\R_6\uparrow}b_{\R_6\downarrow}
        +(b_{\R_0\uparrow} b_{\R_6\downarrow}- b_{\R_0\downarrow}b_{\R_6\uparrow}) \\
        &-(b_{\R_0\uparrow} b_{\R_1\downarrow} - b_{\R_0\downarrow} b_{\R_1\uparrow})
        \big]|\Phi\rangle.
    \end{aligned}
    \label{eq:r1conds}
\end{equation}
These equations reduce to the conditions in the main text when there is a quasilocality site, the $f$ orbitals, that eliminates the double occupancy terms $b_{\R_i\uparrow} b_{\R_i \downarrow}$.
Thus, these conditions are satisfied by states with spin exchange symmetry when there are no doubly occupied loop states, meaning states where clusters of loop states maximize spin remain ground states. There are, however, linearly independent ground states that do not maximize spin due to having a doubly occupied loop state. One such example is

\begin{equation}
    |\Phi'_{\R_0}\rangle = a^\dagger_{\R_0 \uparrow} \sum_{i = 0}^6 a^\dagger_{\R_i \downarrow} |0\rangle,
\end{equation}
a two-particle state consisting of a spin-up loop at the $\R_0$ plaquette surrounded by a superposition of spin-down loops at $\R_0$ and the surrounding plaquettes, as sketched in Fig. \ref{fig:spinexample}. 
It can be verified that the state $|\Phi'_{\R_0}\rangle$ satisfies the three zero-interaction-energy conditions on every site. The conditions in Eq. \eqref{eq:r1conds} are satisfied by $|\Phi'_{\R_0}\rangle$, and the conditions for the remaining sites can be shown to hold as well and follow from simple mappings. For example, the conditions at site $\r_2 = \R_0 + \u_2$ follow from replacing $\R_0 \to \R_1,$ $\R_1 \to \R_0$, and $\R_6 \to \R_2$ in Eq. \eqref{eq:r1conds}.

The two-electron state $|\Phi'_{\R_0}\rangle$ includes a component where the state $\vecg{\varphi}_{\R_0}$ is doubly occupied and has total spin $\langle \Phi'_{\R_0}| S^2_{tot}|\Phi'_{\R_0}\rangle / \langle \Phi'_{\R_0}|\Phi'_{\R_0}\rangle = \frac{6}{7}$. The percolation representation with maximum-spin clusters thus does not hold in the loop basis for $H_K^p$. 
It should be noted that the state $|\Phi'_{\R_0}\rangle$ in Fig. \ref{fig:spinexample}, when written in the orbital basis rather than the loop basis, is equivalent to a two-electron state with one loop state on the central plaquette and a larger loop on the boundary of the six neighboring plaquettes. Thus, in the orbital basis, this state avoids the interaction energy trivially by avoiding doubly occupied sites. However, this state cannot be written as a superposition of maximum-spin clusters of single-plaquette loops, since it has a component where the loop $\R_0$ is doubly occupied, and thus the percolation representation does not hold in terms of the single-plaquette loop basis.

It is important to note that the lack of a percolation representation in the loop state basis does not imply the absence of ferromagnetism in the phase diagram. Indeed, ferromagnetism at flat band half filling can be verified using Mielke's theorem, according to which a system with a lowest- or highest-energy flat band is ferromagnetic at flat band half filling for $U>0$ if and only if the orthogonal projection matrix onto the flat band space is irreducible \cite{Mielke1993, Mielke1999, Tasaki2020fm}. In fact, the lack of a percolation representation in one basis in general does not prevent a percolation representation from being found in another choice of basis, though for the $p$-orbital Hamiltonian, any other basis would likely feature overlaps between more than two states even if quasilocality were satisfied. Thus, while the $p$-orbital system is ferromagnetic at flat band half filling, it is difficult to find a percolation representation with which the onset of ferromagnetism near flat band half filling can be studied, and the most natural basis of loop states $|\psi^{-(p)}_{\R, \sigma}\rangle$ does not admit such a representation.

\section{Particle-Hole Transformation of Total Spin Operator}
\label{appdx:spinph}

We show here for completeness that the particle-hole transformation and total spin operators commute. For notational simplicity, consider electron operators $c_{i \sigma}$, absorbing both the site and orbital indices into the single index $i$. The total spin operator can be written
\begin{equation}
\begin{aligned}
    S^2_{tot} &= (\sum_i \vec{S}_i)^2\\
    &= \frac{1}{4}\sum_{i, j} \sum_{\mu\nu\alpha\beta = \uparrow, \downarrow} c^\dagger_{i\mu}c_{i\nu} c^\dagger_{j\alpha}c_{j\beta} \vecg{\sigma}_{\mu\nu}\cdot\vecg{\sigma}_{\alpha\beta} \\
    &= \frac{1}{4}\sum_{i, j} \sum_{\mu\nu\alpha\beta = \uparrow, \downarrow} (-c_{i\nu}c^\dagger_{i\mu} + \delta_{\mu\nu}) (-c_{j\beta}c^\dagger_{j\alpha} + \delta_{\alpha\beta}) \vecg{\sigma}_{\mu\nu}\cdot\vecg{\sigma}_{\alpha\beta} \\
    &= \frac{1}{4}\sum_{i, j} \sum_{\mu\nu\alpha\beta = \uparrow, \downarrow} c_{i\nu}c^\dagger_{i\mu}c_{j\beta}c^\dagger_{j\alpha} \vecg{\sigma}_{\mu\nu}\cdot\vecg{\sigma}_{\alpha\beta} \\
    &= S^{(ph)2}_{tot}
\end{aligned}
\end{equation}
which follows from the tracelessness of $\sigma^i$ and the fact that $\vecg{\sigma}_{\mu\nu}\cdot\vecg{\sigma}_{\alpha\beta} = \vecg{\sigma}_{\nu\mu}\cdot\vecg{\sigma}_{\beta\alpha}$. The particle-hole transformed spin operator is $\vec{S}^{(ph)}_{tot} = \frac{1}{2}\sum_{i}\sum_{\mu \nu} c_{i\mu} \vecg{\sigma}_{\mu\nu} c^\dagger_{i\nu}$, which is the original spin operator $\vec{S}_{tot}$ with the replacement $c_{i\mu}\leftrightarrow c^\dagger_{i\mu}$. Thus, total spin is preserved by the particle-hole transformation. 

\section{Hund's Coupling Operator Identity}
\label{appdx:Hund}

The Hund's coupling term $H_J$ is a sum of positive semidefinite operators at each honeycomb site,
\begin{equation}
\begin{aligned}
    H_J &= \sum_{\r\in\Lambda_p} h_J(\r),\\
    h_J(\r) &\equiv -J\left(\vec{S}_{\r, p_x} \cdot \vec{S}_{\r, p_y}-\frac{1}{4} n_{\r, p_x} n_{\r, p_y}\right)
\end{aligned}
\end{equation}
where $S^i_{\r, p_{x/y}} = \frac{1}{2}\sum_{\mu, \nu =\uparrow, \downarrow} p^\dagger_{\r,\hat{x}/\hat{y},\mu}\sigma^i_{\mu\nu}p_{\r, \hat{x}/\hat{y}, \nu}$ and $n_{\r, p_{x/y}} = \sum_{\mu = \uparrow, \downarrow} p^\dagger_{\r, \hat{x}/\hat{y}, \mu}p_{\r, \hat{x}/\hat{y},\mu}$. $h_J(\r)$ takes its minimum eigenvalue of $0$ when the $p_x$ and $p_y$ orbitals at site $\r$ are singly occupied and form a spin triplet state, which can be seen explicitly by using the Pauli matrix completeness identity $\sum_{i = 1}^3 \sigma^i_{\alpha\beta}\sigma^i_{\mu\nu} = 2\delta_{\alpha\nu}\delta_{\beta\mu} - \delta_{\alpha\beta}\delta_{\mu\nu}$ to write
\begin{equation}
    \begin{aligned}
    h_J(\r)  = &-\frac{J}{4} \bigg[-n_{\r, p_x} n_{\r, p_y} + \\
    &\sum_{\mu,\nu;\alpha,\beta = \uparrow, \downarrow}p^\dagger_{\r, \hat{x}, \mu}p_{\r, \hat{x}, \nu}p^\dagger_{\r, \hat{y}, \alpha}p_{\r, \hat{y}, \beta}(2\delta_{\alpha\nu}\delta_{\beta\mu} - \delta_{\alpha\beta}\delta_{\mu\nu}) \bigg] \\
    = &\frac{J}{2}\bigg(n_{\r,p_x,\uparrow}n_{\r,p_y,\downarrow} + n_{\r,p_x,\downarrow}n_{\r,p_y,\uparrow} \\
    &- p^\dagger_{\r, \hat{x},\uparrow}p_{\r, \hat{x},\downarrow}p^\dagger_{\r, \hat{y},\downarrow}p_{\r, \hat{y},\uparrow} 
    - p^\dagger_{\r, \hat{x},\downarrow}p_{\r, \hat{x},\uparrow}p^\dagger_{\r, \hat{y},\uparrow}p_{\r, \hat{y},\downarrow} \bigg) \\
    = &-\frac{J}{2}\bigg(\pxup^\dagger\pydown^\dagger\pxup\pydown + \pxdown^\dagger \pyup^\dagger \pxdown\pyup \\
    & - \pxup^\dagger \pydown^\dagger \pxdown \pyup - \pxdown^\dagger \pyup^\dagger \pxup\pydown
    \bigg) \\
    = &\frac{J}{2}(\pydown^\dagger\pxup^\dagger  - \pyup^\dagger\pxdown^\dagger )(\pxup \pydown - \pxdown \pyup) \\
    \equiv &\frac{J}{2} n_{\r, S = 0}.
    \end{aligned}
\end{equation}
The Hund's coupling term can thus be written in terms of a sum of singlet number operators.

\section{Correlated Percolation Simulation Algorithm}
\label{appdx:alg}
In this appendix, we detail the method used for performing Monte Carlo simulations of the correlated percolation problem. The method is similar to that used in Refs. \cite{Maksymenko2012, Liu2019a}. As in the main text, we are considering correlated site percolation in particular, as opposed to bond percolation. We denote the graph $G = (V, E)$ with $V$ the set of vertices and $E$ the set of edges. The general method does not depend on the graph structure, though the graph relevant to the main text is the triangular lattice. The correlated percolation algorithm makes use of many of the subroutines for uncorrelated percolation, particularly generating initial configurations at different fillings and labeling the clusters. 

For uncorrelated percolation, sample configurations can be generated at an exact filling $p$ simply by uniformly selecting sites to fill until $p |V|$ are filled or by filling $p |V|$ sites and performing a random permutation \cite{Newman2001}. These configurations are then already independent samples for uncorrelated percolation that can have clusters labeled by the Hoshen-Kopelman algorithm \cite{Hoshen1976}. Uncorrelated percolation in a range of filling $p_0 \leq p \leq  p_1$ can be efficiently simulated by generating a configuration at filling $p |V| + 1$ from a configuration at filling $p |V|$ and updating the cluster labels using the Newman-Ziff algorithm \cite{Newman2001}. 

For correlated percolation, an uncorrelated percolation configuration can be used as an initial configuration and labeled using the Hoshen-Kopelman algorithm. 
However, as a result of the nonuniform weights, these configurations are not independent samples from the equilibrium distribution and must be moved towards equilibrium by 
applying an update scheme such as the Metropolis-Hastings Monte Carlo algorithm.

The graph and the configuration are stored using the following structures.
\begin{itemize}
    \item \textbf{Neighbor array:} A 2D array of site neighbors of size $|V| \times D_{\text{max}}$, where element $(i, n)$ of the array is the site index $i_n$ for the $n$th neighbor of site $i$ and $D_{\text{max}}$ is the coordination number or the maximum vertex degree in a general graph.
    \item \textbf{Configuration array:} An array $\{l_i\}$ for sites $i=1, \cdots, |V|$ with $l_i = 0$ for empty sites and $l_i$ a positive integer, the \textit{cluster label}, for filled sites.
    \item \textbf{Cluster map:} A map from cluster labels $l$ to \textit{cluster values} $Val(l)$, which are nonzero integers.
    If $Val(l)>0$, $Val(l)$ is the size of the cluster to which sites labeled $l$ belong. If $Val(l)<0$, then $l' \equiv |Val(l)|$ is another cluster label belonging to the same cluster.
    \item \textbf{Proper cluster label:} 
    A label $l^{(k_p)}$ with value $Val(l^{(k_p)}) > 0$ equal to the size of the cluster containing $i$. The proper cluster label $l^{(k_p)}$ is found from the cluster label $l^{(1)} \equiv l_i$ for site $i$ by iteratively evaluating $l^{(k+1)} = |Val(l^{(k)})|$ with $l^{(k_p)}$ the first label in this sequence with positive cluster value.
\end{itemize}
The cluster map scheme is central to the Hoshen-Kopelman algorithm \cite{Hoshen1976}. In the end, the quantities relevant to the percolation transition are the sizes of the clusters rather than the cluster labels, which can be assigned according to any convenient scheme. 
As clusters merge, the Hoshen-Kopelman algorithm keeps track of the new cluster sizes efficiently by 
using negative cluster values to 
avoid the need to update cluster labels for individual sites in the new cluster, and clusters are then identified by the proper cluster labels.

A simpler but less efficient scheme involves traversing all clusters. The cluster values can then all be made positive and equal to the size of the corresponding cluster, with all filled sites in a cluster sharing the same cluster label.
This can be done by starting from an initial unlabeled configuration and scanning through sites on the graph, skipping empty sites as well as previously visited filled sites. 
When an unvisited filled site is reached, it is assigned the smallest available cluster label $l$. 
The cluster containing this site is traversed by iteratively visiting filled neighbors and each site in the cluster is assigned the same cluster label $l$. 
The corresponding $Val(l)$ is set to the size of this cluster. 

The Hoshen-Kopelman algorithm improves efficiency by avoiding the need to fully traverse clusters to relabel filled sites when merging clusters.
This is done by the following using negative cluster values to point to another label in the same cluster. 
When filling a site $i$ would merge multiple clusters to form a larger cluster, the total number of sites $|C_i|$ in the new cluster can be found by consulting the cluster map for the labels of each of the $n_i$ filled neighbors $i_m$ of $i$. 
The smallest proper cluster label $l$ among the filled neighbors is then given the cluster value $Val(l) = |C_i| = 1 + \sum_{m=1}^{n_i}|C_{i_m}|$ and the proper cluster labels of the remaining neighbors $l' > l$ are all given the value $Val(l') = -l$. The merge thus leaves all sites in the cluster with the same proper cluster label $l$ and the need to traverse the entire cluster is avoided at the small cost of one additional cluster map lookup when finding the cluster size in the future.  

Once an initial configuration at filling $p$ is generated and labeled, an equilibrium configuration of the weighted distribution in the canonical ensemble, Eq. \eqref{eq:weights}, is generated by iterating the following procedure. For clarity, we refer directly to the size of the cluster containing site $i$ as $|C_i|$, noting that $|C_i|$ is found by consulting the cluster map for label $l_i$.

\begin{enumerate}
\item Starting from configuration $A$, select one filled site $i$ and one empty site $j$ uniformly at random.
\item Empty site $i$ and calculate the weight ratio $w' \equiv W(A') / W(A)$, where $A'$ is the configuration $A$ with $i$ removed, as follows. 

\begin{enumerate}
    \item Set the initial weight ratio value $w' \leftarrow 1/(|C_i|+1)$ with $C_i \subset A$ the cluster initially containing site $i$. Set a counter $ c \leftarrow 0$ for the number of filled sites traversed.
    \item While the number of unvisited filled neighbors of $i$ is $n_i>0$, pick an unvisited filled neighbor $i_m$ of site $i$ and perform a breadth first search (BFS) of the cluster $C_{i_m}\subset A'$ containing $i_m$.
    
    \begin{enumerate}
        \item During the search, mark visited cluster sites and count the size of the cluster $|C_{i_m}|$.
        \item If during the search all neighbors of $i$ have been visited, break the loop.
        \item If the search completes without visiting all neighbors of $i$, update the weight ratio $w' \leftarrow w' \times(|C_{i_m}| + 1)$ and the counter $c \leftarrow c + |C_{i_m}|$.
    \end{enumerate}
    
    \item The size of the remaining cluster must be $|C_i| - 1 - c$, and the weight is updated accordingly, $w' \leftarrow w'\times(|C_i| - c)$.
\end{enumerate}

\item Fill site $j$ and calculate the weight ratio $w'' \equiv W(A'')/W(A')$, where $A''$ is the configuration $A'$ with $j$ filled, by consulting the cluster map for $A'$ as follows.

\begin{enumerate}
    \item Find the proper cluster labels for each filled neighbor of $j$ and create a list of $n_c$ filled neighbors $j_m$ with one $j_m$ for each distinct proper cluster label. Note that $n_c$ is the number of clusters neighboring $j$ in $A'$.
    \item Store the minimum proper cluster label $l_{min}$ among proper cluster labels $l_m$ for the $j_m$. Set $Val(l_{min}) \leftarrow 1 + \sum_{m=1}^{n_c} |C_{j_m}|$. For all $l_m > l_{min}$, set $Val(l_m) \leftarrow -l_{min}$.
    \item Set the weight ratio $w'' \leftarrow (1+\sum_{m=1}^{n_c} |C_{j_m}|)/\prod_{m=1}^{n_c} (|C_{j_m}| +1)$.
\end{enumerate}

\item Accept the trial configuration $A''$ if $w \equiv w'\times w'' = W(A'')/W(A) >1$. Otherwise accept $A''$ with probability $w$.
\end{enumerate}

Explicitly filling site $j$ and updating the cluster labels can be delayed until after the updated configuration is accepted, since the relative weight $w''$ for filling a site can be calculated from the cluster map before filling the site.
When emptying a site $i$, is it unfortunately necessary in the worst case to traverse all but one of the clusters $C_{i_n}$ neighboring $i$, since connectivity cannot be determined by examining only the neighbors of $i$. Since each cluster $C_{i_n}$ must be traversed anyway when emptying site $i$, it is reasonable to assign a new cluster label to each traversed site of the cluster to minimize the number of negative cluster values in the cluster map. We allow early termination in the update step when all neighbors of $i$ have been visited, make use of the fact that knowing the size of the original cluster $C_i$ with $i$ filled means one of the clusters $C_{i_n}$ with $i$ empty need not be explicitly traversed. 
Performing a breadth first search of the clusters maximizes the chances of early termination occurring before the last cluster is traversed. As opposed to filling a site, emptying a site is performed explicitly to generate a trial configuration, since all but one of the neighboring clusters must be traversed to update the cluster labels. If the configuration is rejected, the site can simply be filled in.

Equilibrium configurations in the grand canonical ensemble at fixed fugacity $z$, with weights given by Eq. \eqref{eq:gcweights}, follow from an initial configuration by selecting a site uniformly at random and attempting to empty it if filled or fill it if empty, using the corresponding subroutine from the canonical ensemble simulations. The initial configuration can be randomly generated at any filling or even taken to be empty.


%

\end{document}